\documentclass[traditabstract,bibyear]{aa} % for a printer version

\newcommand{\gtsim}{\protect\raisebox{-0.5ex}{$\:\stackrel{\textstyle >}
        {\sim}\:$}}
\newcommand{\ltsim}{\protect\raisebox{-0.5ex}{$\:\stackrel{\textstyle <}
        {\sim}\:$}}

\usepackage{color}
\usepackage{graphicx}
\usepackage{natbib,twoopt}
\bibpunct{(}{)}{;}{a}{}{,} % to follow the A&A style
\makeatother

\begin{document}

\title{X-shooter spectroscopy of young stellar objects in Lupus:}
\subtitle{Lithium, iron, and barium elemental abundances\thanks{Based on observations collected at the European Organization 
for Astronomical Research in the Southern Hemisphere (Paranal, Chile) under programs 084.C-0269(A), 085.C-0238(A), 085.C-0764(A), 
086.C- 0173(A), 087.C-0244(A), 089.C-0143(A), 093.C-0506(A), 095.C-0134(A), and 097.C-0349(A).}\fnmsep
\thanks{This paper is dedicated to the memory of Prof. Francesco Palla, who passed away in 2016.}
}
   
\author{
K. Biazzo \inst{1} \and A. Frasca \inst{1} \and J. M. Alcal\'a \inst{2} \and M. Zusi \inst{3} \and  E. Covino \inst{2} \and  S. Randich \inst{4} 
\and  M. Esposito \inst{2} \and C. F. Manara \inst{5} \and  S. Antoniucci \inst{6} \and  B. Nisini \inst{6} \and E. Rigliaco \inst{7} \and F. Getman \inst{2}}

\offprints{K. Biazzo}
\mail{katia.biazzo@oact.inaf.it}

\institute{INAF - Osservatorio Astrofisico di Catania, Via S. Sofia 78, I-95123 Catania, Italy
\and INAF - Osservatorio Astronomico di Capodimonte, Salita Moiariello 16, I-80131 Napoli, Italy
\and INAF - Istituto di Astrofisica e Planetologia Spaziali, via del Fosso del Cavaliere 100, I-00133 Rome, Italy
\and INAF - Osservatorio Astrofisico di Arcetri, Largo E. Fermi 5, I-50125 Firenze, Italy
\and Scientific Support Office, Directorate of Science, European Space Research and Technology Centre (ESA/ESTEC), Keplerlaan 1, 2201 AZ Noordwijk, The Netherlands
\and INAF - Osservatorio Astronomico di Roma, via Frascati 33, I-00078 Monte Porzio Catone, Italy
\and INAF - Osservatorio Astronomico di Padova, vicolo dell'Osservatorio 5, I-35122 Padova, Italy
}

\date{Received .../ accepted ...}

\abstract
{}
{With the purpose of performing a homogeneous determination of elemental abundances for members of the Lupus T association, we analyzed 
three chemical elements: lithium, iron, and barium. The aims were: 1) to derive the lithium abundance for the almost complete sample ($\sim 90\%$) 
of known class II stars in the Lupus I, II, III, and IV clouds; 2) to perform chemical tagging of a region where few iron abundance 
measurements have been obtained in the past, and no determination of the barium content has been done up to now. We also investigated 
possible barium enhancement at the very young age of the region, as this element has become increasingly interesting in the last few years 
following the evidence of barium over-abundance in young clusters, the origin of which is still unknown.}
{Using the X-shooter spectrograph mounted on the Unit 2 (UT2) at the Very Large Telescope (VLT), we analyzed the spectra of 89 cluster members, both class II (82) and class III (7) stars. 
We measured the strength of the lithium line at $\lambda$6707.8\,\AA\,and derived the abundance of this element through equivalent width measurements 
and curves of growth. For six class II stars we also derived the iron and barium abundances using the spectral synthesis method and the code MOOG. 
The veiling contribution was taken into account in the abundance analysis for all three elements.}
{We find a dispersion in the strength of the lithium line at low effective temperatures and identify three targets with severe Li depletion. The 
nuclear age inferred for these highly lithium-depleted stars is around $15$\,Myr, which exceeds by an order of magnitude the isochronal one. We 
derive a nearly solar metallicity for the members whose spectra could be analyzed. We find that Ba is over-abundant by $\sim 0.7$\,dex with respect 
to the Sun. Since current theoretical models cannot reproduce this abundance pattern, we investigated whether this unusually large Ba content 
might be related to effects due to stellar parameters, stellar activity, and accretion. }
{We are unable to firmly assess whether the dispersion in the lithium content we observe is a consequence of an age spread. As in other star-forming 
regions, no metal-rich members are found in Lupus, giving support to a recent hypothesis that the iron abundance distribution of most of the nearby 
young regions could be the result of a common and widespread star formation episode involving the Galactic thin disk. Among the possible
causes or sources for Ba enhancement examined here, none is sufficient to account for the over-abundance of this element at a $\sim 0.7$\,dex level..
}  
	   
\titlerunning{Lithium, Iron, and Barium abundances in Lupus YSOs}
\authorrunning{K. Biazzo et al.}
\maketitle

\section{Introduction}
\label{sec:intro}

The determination of elemental abundances in nearby ($<$~$500$ pc) star-forming regions (SFRs) is important for a variety of astrophysical problems, 
in both exo-planetary and stellar contexts. The members of these regions are still close to their birthplace. Their elemental 
abundances are thus fundamental to trace the present chemical pattern of the Galactic thin disk in the solar neighborhood 
and the interstellar medium in which they are immersed. 

In the last three decades, an increasing number of studies have focused on the abundance measurements of iron and other elements in SFRs, young clusters, 
and associations, and specifically in their low-mass members (e.g., \citealt{padgett1996, cunhaetal1998, jamesetal2006, santosetal2008, 
gonzalez-hernandez2008, vianaalmeidaetal2009, dorazirandich2009, biazzoetal2011a, biazzoetal2011b, dorazietal2011, 
taberneroetal2012, biazzoetal2012a, biazzoetal2012b, spinaetal2014a, spinaetal2014b}). In particular, three elements have gradually been more 
analyzed in studies investigating the chemical pattern of young regions, namely lithium (Li), iron (Fe), and barium (Ba).

Lithium is a fragile element, which is burnt at temperatures of $\sim 3 \times 10^6$\,K (see, e.g., \citealt{bildstenetal1997}). 
Temperatures of this order can be easily reached in the interior of a low-mass pre-main sequence (PMS) star as it contracts towards the main 
sequence (\citealt{bodenheimer1965}). Therefore, low-mass stars ($\sim 0.05-0.7\,M_\odot$) deplete their initial Li content during the PMS phase; 
the depletion timescale depends on stellar mass, with stars of $\sim 0.7-0.2\,M_\odot$ starting to burn it after $\sim 2-15$\,Myr and stars of 
$< 0.2\,M_\odot$ after $\sim 20-25$ Myr (see \citealt{baraffeetal2015}, and references therein). Therefore, lithium abundance has been widely used 
as an independent and reliable method to estimate the ages of low-mass members of young regions (see, e.g., 
\citealt{songetal2002, whitehillenbrand2005, pallaetal2005, pallaetal2007, saccoetal2007, yeejensen2010, sergisonetal2013, limetal2016}).

The determination of the abundance of iron, iron-peak, and alpha elements is important in the field of star 
formation. In fact, several studies have provided hints that regions where star formation has ceased generally share a metallicity close to 
the solar value, while SFRs where the molecular gas is still present seem to be characterized by a slightly lower iron content 
(\citealt{biazzoetal2011a, spinaetal2014b}, and references therein). Whether this is due to low-number statistics (both in regions and in number 
of analyzed targets per region) or inhomogeneous and uncertain methodologies is still debated. \cite{spinaetal2014b} recently 
claimed that the metal-poor nature of these young environments could be the result of a common and widespread star formation episode involving 
the Gould Belt, and giving birth to most of the SFRs, and stars in the solar neighborhood 
(see also \citealt{guilloutetal1998}). Ho\-we\-ver, the possibility of a more complex process of chemical evolution 
that involved a much larger area in the disk of the Milky Way Galaxy is not excluded (\citealt{spinaetal2017}).

Barium is synthesized by neutron capture reactions mostly by the so called $s$-process occurring in asymptotic giant branch (AGB) stars and represents 
an excellent tracer of chemical enrichment mechanisms in the Galaxy; for this reason, many studies have been focused on deriving barium abundance in 
both field stars in the halo and thick/thin disk, and in stellar clusters older than $\sim$100 Myr. Fewer studies are available for young clusters and 
associations, while to our knowledge no determination of barium abundance for late-type members of SFRs exists to date. \cite{dorazietal2009} detected a 
trend of increasing [Ba/Fe] with decreasing age from roughly solar abundance up to $\sim 0.3$\,dex for open clusters in the age range $\sim 0.5-4.5$\,Gyr. 
\cite{dorazietal2009} reproduced such behavior by assuming higher Ba yields from low-mass AGB stars in their Galactic chemical evolution models (see also 
\citealt{maiorcaetal2014}). A further increase in still younger clusters, up to $\sim 0.6-0.7$\,dex at $\sim 35$\,Myr, was also found by \cite{dorazietal2009}; 
this behavior is not reproduced by the same models. They argued that a process creating Ba in the last dozen Myr through Galactic chemical evolution is 
quite unlikely, unless local enrichment is invoked. Other recent studies confirmed the presence of high Ba abundance in $\sim 30-50$\,Myr old stars 
(\citealt{desideraetal2011, dorazietal2012}). Several hypotheses (high level of chromospheric activity, uncertainty in stellar parameters, effects 
of the stratification in temperature of the model atmosphere, non local thermodynamic equilibrium, NLTE, corrections) were proposed to reproduce the high 
Ba content in young clusters, but all of them failed in explaining the observed over-abundance.

Here, we present a systematic and homogeneous ana\-ly\-sis of elemental abundances of PMS stars in the Lupus cloud complex. 
Lupus is one of the most nearby ($d \sim 150-200$ pc) and largest low-mass star-forming regions (see \cite{comeron2008} for a review). 
Similarly to other regions (e.g., Taurus, Chamaeleon, Ophiucus, Corona Australis), a large variety of objects in various stages of evolution are present 
in Lupus. The region of the sky occupied by the Lupus clouds is almost devoid of early-type stars and shows no sign of ongoing high-mass star forming 
activity, although the large number of OB members of the Scorpius-Centaurus SFR in the vicinity of the Lupus complex implies 
the existence of an ambient field of high energy sources, which are likely to have played an important role in the evolution and possibly the origin 
of the Lupus complex (\citealt{tachiharaetal2001}). This makes the Lupus stellar population of particular interest for comparative studies with 
other nearby regions, such as Taurus-Auriga and Chamaeleon, which have a similar mass of molecular gas and low-mass star formation activity, but which evolved 
more in isolation and relatively unperturbed (\citealt{comeron2008}).

This is a companion paper to previous studies focused on the in\-ve\-sti\-ga\-tion of the accretion properties of Lupus PMS stars 
(\citealt{alcalaetal2014, alcalaetal2017}), as well as the determination of their stellar parameters and acti\-vi\-ty indicators (\citealt{frascaetal2017}). 
We used the same data as in those works, namely spectra acquired with the X-shooter spectrograph on the Very Large Telescope (VLT; Paranal, Chile), in order 
to study the abundance of lithium, iron, and barium of the low-mass ($\sim 0.025-1.8\,M_\odot$) PMS stars in Lupus. While a few studies of elemental 
abundance of iron, silicon, and nickel in a handful of class III stars in this region have been performed in the past (see, \citealt{santosetal2008}, 
and references therein), and only one class II was analyzed in terms of iron abundance (\citealt{padgett1996}), a homogeneous and self-consistent 
analysis of Li, Fe, and Ba in a significant sample of the class II sources is still lacking. In this paper, we aim to cover this gap.

The outline of this paper is as follows. The data sample is presented in Sect.\,\ref{sec:obs}. In Sect.\,\ref{sec:elem_abun}, 
we derive the elemental abundances of lithium, iron, and barium. We then discuss the implications 
of our findings in Sect.\,\ref{sec:results_discussion}. In Sect.\,\ref{sec:conclusions} a summary of our results is presented. 

\section{Data set}
\label{sec:obs}
The data set for this paper is exactly the same as in our  previous works (\citealt{alcalaetal2014, alcalaetal2017, frascaetal2017}). 
All the criteria for the target selection, as well as observational strategy, are provided in those papers. We adopt the same 
listing order as in \cite{frascaetal2017} in our Table~\ref{tab:param}.

The final sample includes 82 class II and seven class III sources in Lupus clouds I, II, III, and IV. Forty-three objects were observed in 
2010-2012 during the INAF ({\it Istituto Nazionale di Astrofisica}) guaranteed time observations (GTO; \citealt{alcalaetal2011, alcalaetal2014}), 
forty sources were observed in 2015 and 2016 during the ESO (European Southern Observatory) periods 95 and 97, and the remaining 
six were taken from the ESO archive (see \citealt{alcalaetal2017, frascaetal2017}). All targets were observed using the 0\farcs{9} slit at a resolution 
of $R\sim8\,800$ in the VIS (visible) arm, with the exception of four (Sz\,68, Sz\,74, Sz\,83, and Sz\,102) in P95 and P97, and those of the ESO archive, which were 
observed using the 0\farcs{4} VIS slit at a resolving power of $R\sim17\,400$. The sample comprises 82 class II and seven class III sources. For details about 
data reduction, observing log-book, and selection criteria, we refer to \cite{alcalaetal2014,alcalaetal2017} and \cite{manaraetal2013}. Throughout this 
paper, stellar parameters (effective temperature $T_{\rm eff}$, surface gra\-vi\-ty $\log g$, projected rotational velocity $v \sin i$, veiling $r$) and 
membership information (radial velocity $v_{\rm rad}$) were taken from \cite{frascaetal2017}, while spectral types were taken from \cite{alcalaetal2017} and 
\cite{manaraetal2013}. The sample of class II objects in the aforementioned Lupus clouds is complete at more than $\sim 90$\% level 
(\citealt{alcalaetal2017}). 

\section{Analysis}
\label{sec:elem_abun}
Lithium abundances were measured from the line equi\-va\-lent 
widths and by using appropriate curves of growth (Sect.\,\ref{sec:lithium_abun}), while iron and barium abundances were determined 
through the spectral synthesis method and the {\it synth} driver within the MOOG code (\citealt{sneden1973}; see Sect.\,\ref{sec:iron_ba_abun}). 
This is the most adequate strategy to derive the abundances, given the relatively low resolution of our spectra.

\subsection{The lithium line}
\subsubsection{Equivalent widths}
\label{sec:lithium_abun}
Equivalent widths of the lithium line ($EW_{\rm Li}$) at $\lambda=6707.8$\,\AA\, were measured by direct integration or by Gaussian fit using the 
IRAF\footnote{IRAF, Image Reduction and Analysis Facility, is distributed by the National Optical  Astronomy Observatory, which is operated by the 
Association of the Universities for Research in Astronomy, Inc. (AURA) under cooperative agreement with the National Science Foundation.} task {\sc splot}. 
At the X-shooter re\-so\-lu\-tion the \ion{Li}{i} line is blended with the \ion{Fe}{i} $\lambda$ 6707.4\,\AA~line, whose contribution was 
subtracted using the empirical correction between $EW_{\rm Fe}$ and the $B-V$ color by \cite{soderblometal1993}. 
To estimate the correction, we considered the $T_{\rm eff}$ determinations by \cite{frascaetal2017} and the calibration by 
\cite{pecautmamajek2013}. For stars cooler than 4000 K, the real continuum is no longer visible, due to the increasing strength of molecular 
bands. Therefore, the lithium equivalent widths are referred to the local pseudo-continuum, and the molecular bands are 
the major blending sources. However, we integrated the lithium line including the blends, as in \cite{pallaetal2007} to be consistent with 
their curves of growth (see Sect.\,\ref{sec:lithium_abun}).

Errors in $EW_{\rm Li}$ were evaluated by determining the signal-to-noise ($S/N$) ratio at wavelengths adjacent 
to the lithium line and by multiplying its reciprocal by the width 
of the integration range. Typical errors in lithium equivalent widths are of 20-25\,m\AA\,(see Table\,\ref{tab:param}). 

\begin{figure}[t!]
\begin{center}
\includegraphics[width=9cm]{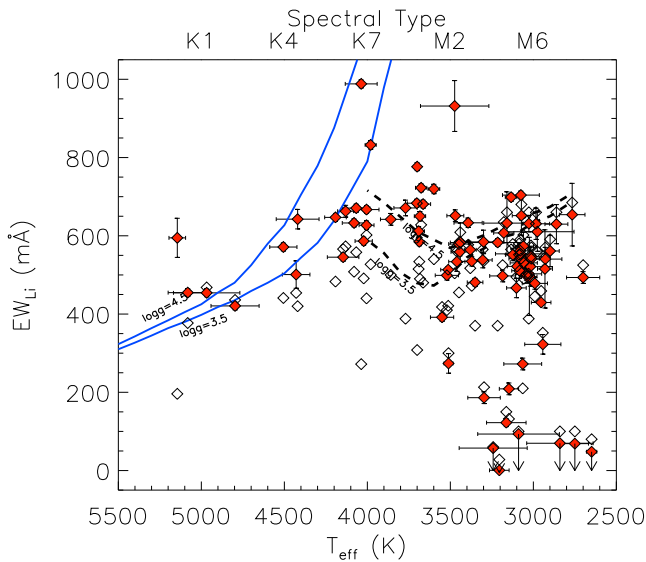}
\caption{Lithium equivalent width versus effective temperature for the studied sample. Open symbols refer to measured $EW_{\rm Li}$, while 
filled red symbols represent the lithium equivalent widths after correction for blending with the iron line and spectral 
veiling. Overplotted as black dashed and blue solid lines are the COGs at $\log g=3.5, 4.5$ and lithium abundance of 3.5~dex from \cite{pallaetal2007} 
and \cite{pavlenkomagazzu1996}, respectively. Arrows indicate upper limits (see text). Spectral types as in \cite{luhmanetal2003} for M-type young 
stellar objects (YSOs) and \cite{kenyonhartmann1995} for K-type YSOs are also marked (see also \citealt{alcalaetal2017} and 
Table\,\ref{tab:param}).}
\label{fig:ewli_teff} 
\end{center}
\end{figure}

\begin{figure}[t!]
\begin{center}
\includegraphics[width=9cm]{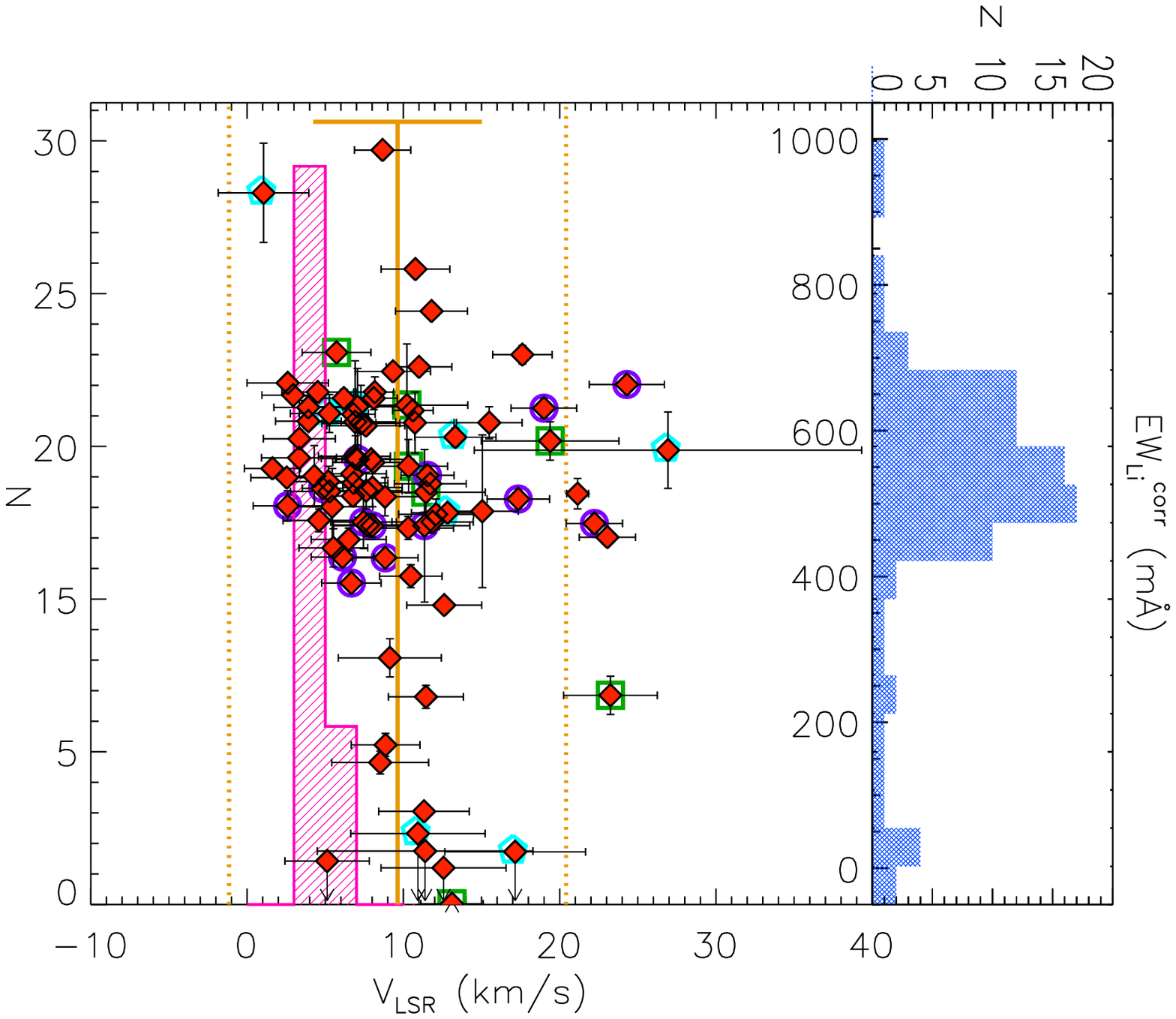}
\caption{Corrected lithium equivalent width versus radial ve\-lo\-ci\-ty (from \citealt{frascaetal2017}) in the LSR. Different symbols are used for class III 
(open green squares), transitional disk (violet circles), and sub-luminous or flat spectral energy distribution objects (cyano pentagons), as defined in 
\cite{alcalaetal2017}. The vertical solid line indicates the mean $V_{\rm LSR}$ ($9.6\pm5.4$ km/s) of our targets, while the dotted ones represent the 
$<V_{\rm LSR}> \pm 2\sigma$ va\-lues. Upper limits are marked by arrows. The magenta hatched histogram in the background of the central panel represents the 
velocity distribution of gas condensations in Lupus as derived by \cite{vilas-boasetal2000}. The histogram on the right panel shows our $EW_{\rm Li}$ 
distribution, with a peak around $\sim 560$\,m\AA.}
\label{fig:ewli_vrad} 
\end{center}
\end{figure}

Our spectra are affected by spectral veiling ($r$), that is, the amount of continuum excess emission, which fills in the lines. 
Measured $EW_{\rm Li}$ were corrected for this contribution using the $r$ values expressed in units of the photospheric continuum and determined by 
\cite{frascaetal2017} in several spectral regions. We applied the relationship $EW_{\rm Li}^{\rm corr}=EW_{\rm Li}(1 + r)$, where 
$EW_{\rm Li}^{\rm corr}$ is the corrected value for the lithium equivalent width. For the \ion{Li}{i} 
line at $\lambda=6707.8$\,\AA,\,we adopted the mean value ($r_{6700}$) between $\lambda=6200$\,\AA\,and $\lambda=7100$\,\AA\,(see Table\,\ref{tab:param}). 
As in \cite{frascaetal2017}, all values of veiling $\le 0.2$ are considered non detectable and set equal to zero. This is the case of about 65\% 
of the targets, while, for the rest of the sample, $r_{6700}$ ranges from $\sim 0.3$ up to $\sim 3$. Figure~\ref{fig:ewli_teff} shows the 
$EW_{\rm Li}$ versus $T_{\rm eff}$ plot for the 89 Lupus members listed in Table~\ref{tab:param}, as obtained from our measurements and after the correction 
both for the blending with the iron line and the veiling contribution as explained above. The distribution of the corrected equivalent width peaks at 
$\sim 560$\,m\AA\,(see also Fig.\,\ref{fig:ewli_vrad}), with most of the spread at a given $T_{\rm eff}$ reduced after the aforementioned corrections. The 
residual scatter in $EW_{\rm Li}^{\rm corr}$ could be due to measurement errors, but we cannot exclude the possibility that part of the dispersion may be due to stars with 
Li depletion. For five targets (Lup\,706, Par-Lup3-4, 2MASS\,J16085953-3856275, SSTc2d\,J154508.9-341734, and 2MASS\,J16085373-3914367) 
we were not able to measure $EW_{\rm Li}$ because of low $S/N$ and/or high veiling contribution. However, the lithium content of these objects 
is below the level of other stars with similar effective temperature. For one target (Sz\,94) we did not detect the line (see 
Table\,\ref{tab:param} and discussion in Sect.\,\ref{sec:lithium_depl}). We consider these six targets as upper limits, since the error in 
equivalent width is larger than the $EW_{\rm Li}$ value. The rest of the targets with low lithium content ($EW_{\rm Li}^{\rm corr} \ltsim 200$\,m\AA) have 
negligible veiling ($r_{6700} \ltsim 0.2$) and suggest a possible large amount of Li depletion. They will be discussed in Sect.\,\ref{sec:lithium_depl}.

Figure \ref{fig:ewli_vrad} shows the position of all members in an $EW_{\rm Li}^{\rm corr}$ versus radial velocity (RV) plot, where RVs by \cite{frascaetal2017} 
were transformed to the local standard of rest (LSR). In the same plot, the RV distribution of the gas derived by \cite{vilas-boasetal2000} from the 
$^{13}$CO ($J = 1 - 0$) transition in 35 dense molecular cores in Lupus I, II, III, and IV is shown. Our RV distribution is in good agreement, within the 
errors, with the average velocity of the gas, that is, $<V_{\rm LSR}>_{\rm gas}=4.7\pm0.3$\,km/s, as also found in other regions 
(\citealt{biazzoetal2012a,darioetal2017}). All but six stars (Sz\,66, Sz\,91, SSTc2d160901.4-392512, Sz\,123A, 
Sz\,102, and Sz\,122) are confined inside $\pm 2\sigma$ from the peak at 9.6 km/s of the RV distribution, where $\sigma = 5.4$ km/s. The six stars outside 
the $<V_{\rm LSR}> \pm 2\sigma$ distribution could be spectroscopic binaries because both their high lithium content and proper motions 
(\citealt{girardetal2011,lopezmartietal2011}) strongly suggest membership. Five of them have large $EW_{\rm Li}^{\rm corr}$, 
which gives support to their membership to the Lupus SFR. One of these, Sz\,102, shows a large $V_{\rm LSR}$ error, maybe due to the 
high $v \sin i$ and veiling (see \citealt{frascaetal2017}). The star with low $EW_{\rm Li}^{\rm corr}$ and outside the 
$<V_{\rm LSR}> \pm 2\sigma$ distribution is Sz\,122, a class III object with $v \sin i \sim 150$\,km/s. This high-rotational velocity 
target is suspected to be a spectroscopic binary, as reported by \cite{stelzeretal2013}. 

\subsubsection{Elemental abundances} 
\label{sec:lithium_abun}
\begin{figure}[t!]
\begin{center}
\includegraphics[width=9cm]{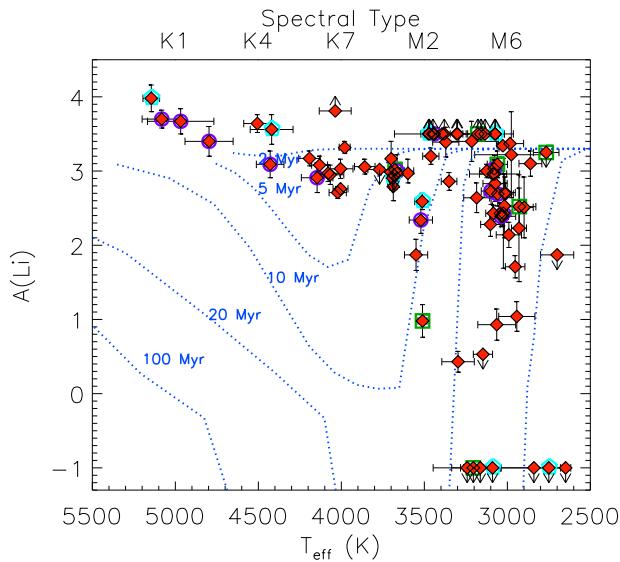}
\caption{Lithium abundance versus effective temperature. The ``lithium isochrones'' by \cite{baraffeetal2015} in the 
2--100 Myr range are overlaid with dotted lines. Arrows refer to upper and lower limits. 
% typically due to low $S/N$ of the spectra or small $EW_{\rm Li}$ values and validity limits of the COGs, respectively. 
Open squares, circles, and pentagons represent the position of the class III, transitional disks, and sub-luminous or flat energy distribution 
targets, respectively. Spectral types as in Fig.\,\ref{fig:ewli_teff} are also shown.}
\label{fig:AbunLi_Teff} 
\end{center}
\end{figure}

\begin{figure}[t!]
\begin{center}
\includegraphics[width=9cm]{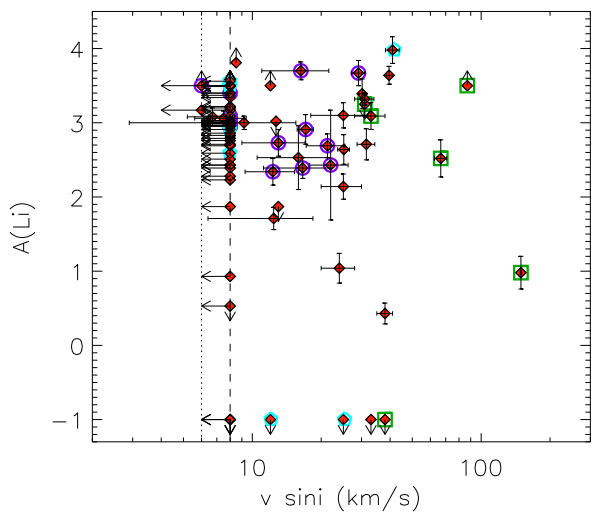}
\caption{Lithium abundance versus projected rotational velocity. Symbols as in Fig.\,\ref{fig:AbunLi_Teff}. The dashed and dotted lines represent 
the upper limits at 8 and 6 km/s of our VIS spectra acquired with the slit width at 0\farcs9 and 0\farcs4, respectively (see \citealt{frascaetal2017}).}
\label{fig:AbunLi_vsini} 
\end{center}
\end{figure}

Lithium abundances, $A$(Li), were estimated from the $EW_{\rm Li}^{\rm corr}$ measured in this work, the atmospheric pa\-ra\-me\-ters 
($T_{\rm eff}$, $\log g$) taken from \cite{frascaetal2017}, and using the NLTE curves of growth (COGs) reported by 
\cite{pavlenkomagazzu1996} for $T_{\rm eff}>4000$~K, the COGs by \cite{pallaetal2007} for $T_{\rm eff}<3500$~K, and the average of these COGs for 
$3500<T_{\rm eff}<4000$~K. The main source of error in $A{\rm (Li)}$ comes from the uncertainty in stellar parameters ($T_{\rm eff}$ and $\log g$), listed 
in \cite{frascaetal2017}, in our measurements of lithium equivalent widths, and from the \cite{soderblometal1993} relationship. We therefore took 
into account all these four error sources and estimated the error in $A{\rm (Li)}$ by adding them in quadrature. The global uncertainties 
range from $\sim 0.1-0.3$~dex up to $\sim 0.4-0.7$~dex for cool stars ($T_{\rm eff}\sim 2800-3000$~K), and are around $0.1-0.2$~dex for 
targets at $\sim 5000$\,K. Finally, uncertainty of $\sim 0.1$ in $r_{6700}$ translates into errors around $\sim 0.1$\,dex and $\sim 0.3$\,dex 
in $A{\rm (Li)}$ for warm and cool stars, respectively. In Figs.~\ref{fig:AbunLi_Teff} and \ref{fig:AbunLi_vsini} we show the lithium abundance as a function 
of the effective temperature and the rotational velocity, respectively (see also Table~\ref{tab:param}). Upper and lower limits in $A{\rm (Li)}$ result from 
the range of validity of the COGs as a function of $T_{\rm eff}$ and $\log g$. Furthermore, upper limits in $EW_{\rm Li}$ translate also into upper limits in 
$A{\rm (Li)}$. Most of the stars have Li abundances between $A{\rm (Li)} \sim 2$ and 4 dex, with a peak at around 3.1 dex, 
independently of their classification in class II, class III, transitional disks, or sub-luminous (see \citealt{alcalaetal2017}, 
and their references therein, for the classification). A spread of Li abundance appears for stars cooler than about 3500\,K, regardless of the 
uncertainties and upper or lower limits. As shown in Fig.\,\ref{fig:AbunLi_vsini}, the scatter cannot be ascribed to a spread in projected rotational 
velocity as the stars with low-Li content ($A{\rm (Li)} \ltsim 2$ dex) have $v \sin i$ spanning from $\sim 40$\,km/s down to less than 8\,km/s. 
The only exception is Sz\,122, which we do not consider in the analysis because it may be an unresolved spectroscopic binary 
(see \citealt{stelzeretal2013}). The lack of $A{\rm (Li)}$-$v \sin i$ connection is in line with the fact that, at the young ages of our 
targets, there is no evidence for lithium-rich fast rotators and lithium-depleted slow rotators; Li abundance appears to be poorly affected by 
rotationally induced mixing arising from angular momentum loss (see \citealt{bouvieretal2016} for a recent discussion about the lithium-rotation connection 
at very young ages). Moreover, since all members are supposed to have formed from the same molecular cloud, we expect our analysis to be free from significant 
star-to-star differences in the initial chemical composition. Therefore, we do not exclude {\it a priori} the possibility that some of the late-type 
stars underwent lithium depletion. This issue will be discussed in Section~\ref{sec:lithium_depl}.\\

\subsection{Spectral synthesis of Fe and Ba lines}
\label{sec:iron_ba_abun}
The spectral synthesis for deriving iron and barium abundances was carried out by employing the code MOOG (\citealt{sneden1973}; 2013 version) and the 
\cite{kurucz1993} set of model atmospheres. We then adopted the 
stellar parameters ($T_{\rm eff}$, $\log g$, $v \sin i$) determined by \cite{frascaetal2017} and con\-si\-de\-ring the stars with: $i)$ $T_{\rm eff}>4400$\,K, 
to avoid the contribution of molecular bands, which would severely affect the determination of [Fe/H] through their uncertain opacities (see, e.g., 
Appendix B in \citealt{biazzoetal2011a}); $ii)$ low vei\-ling ($r < 1.0$), as it reduces the depth of absorption lines, introducing a further uncertainty 
in [Fe/H] determinations; $iii)$ low $v \sin i$ ($< 40$ km/s) to avoid severe line blending. In the end, six targets fulfill these requirements 
(SSTc2d\,J160830.7-382827, RY\,Lup, MY\,Lup, Sz\,68, Sz\,133, and SSTc2d\,J160836.2-392302). SSTc2d\,J160830.7-382827 and MY\,Lup are weak 
accretors with transitional disks, RY\,Lup has a transitional disk, Sz\,68 is a weak accretor, Sz\,133 is a sub-luminous object, 
and SSTc2d\,J160836.2-392302 has a most probable transitional disk (see \citealt{ansdelletal2016, alcalaetal2017} for details). 

\subsubsection{Iron abundance}
\label{sec:iron_abun}
Despite the extremely wide wavelength coverage, the re\-la\-ti\-ve\-ly low resolution of the X-shooter spectra prevented us from using a very wide 
spectral range with unblended and isolated iron lines. For this reason, we chose to derive the iron abundance using the spectral synthesis method in a wavelength 
window of $\sim 20$\,\AA\,around the \ion{Li}{i} line at 6707.8\,\AA. This spectral range proved to be very suitable for reliable iron 
abundance measurements (see \citealt{dorazietal2011}, and references therein, for details). We con\-si\-de\-red the line list employed in \cite{dorazietal2011}, 
with carefully derived atomic parameters. As solar iron abundance we considered the value of \cite{asplundetal2009}, that is, 
$\log n({\rm Fe})_\odot = 7.50 \pm 0.04$\,dex. We refer to \cite{dorazietal2011}, and references therein, for detailed explanations of the method. 
 
\begin{figure*}[t!]
\begin{center}
\includegraphics[width=5.5cm]{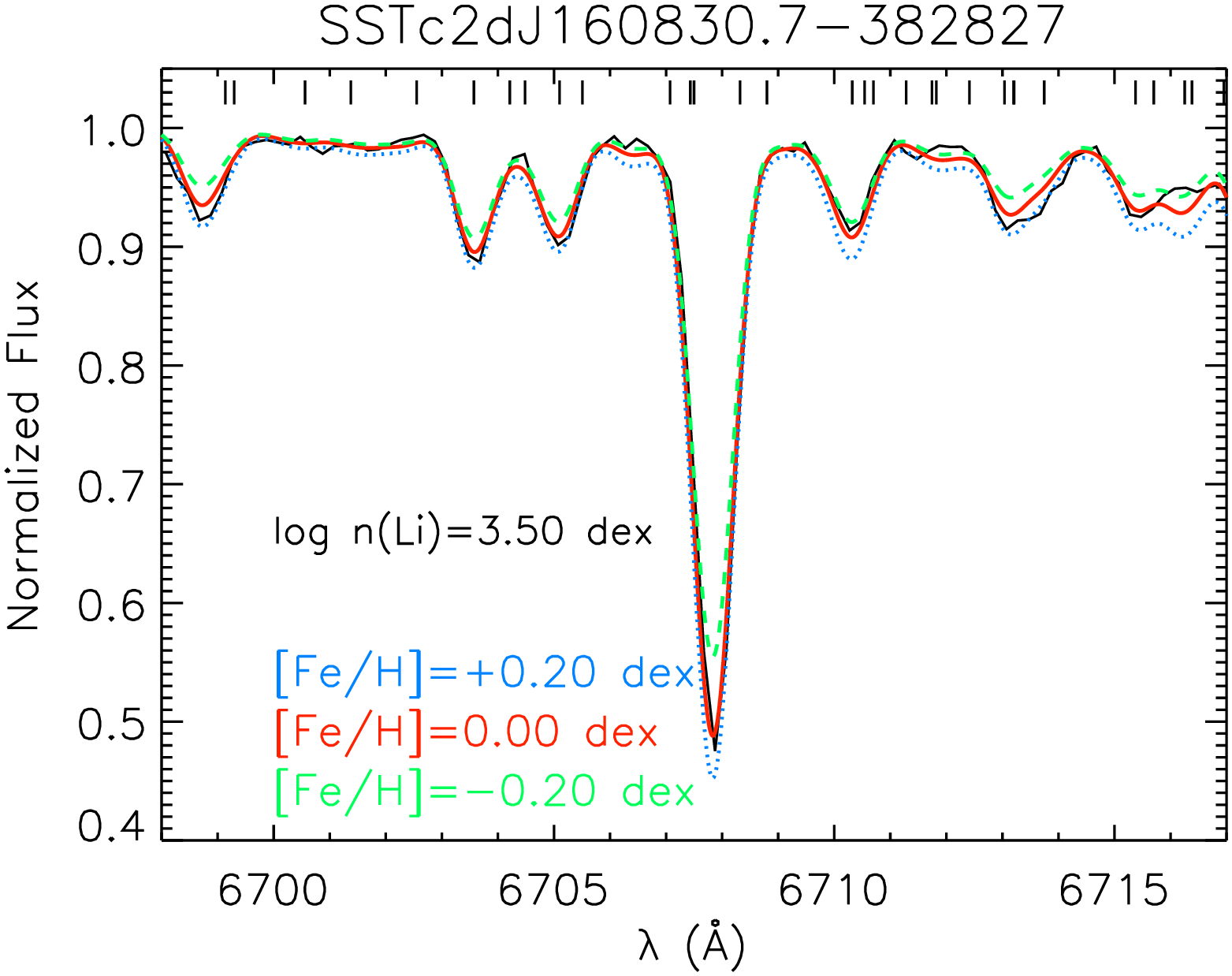}
\hspace{.5cm}
\includegraphics[width=5.5cm]{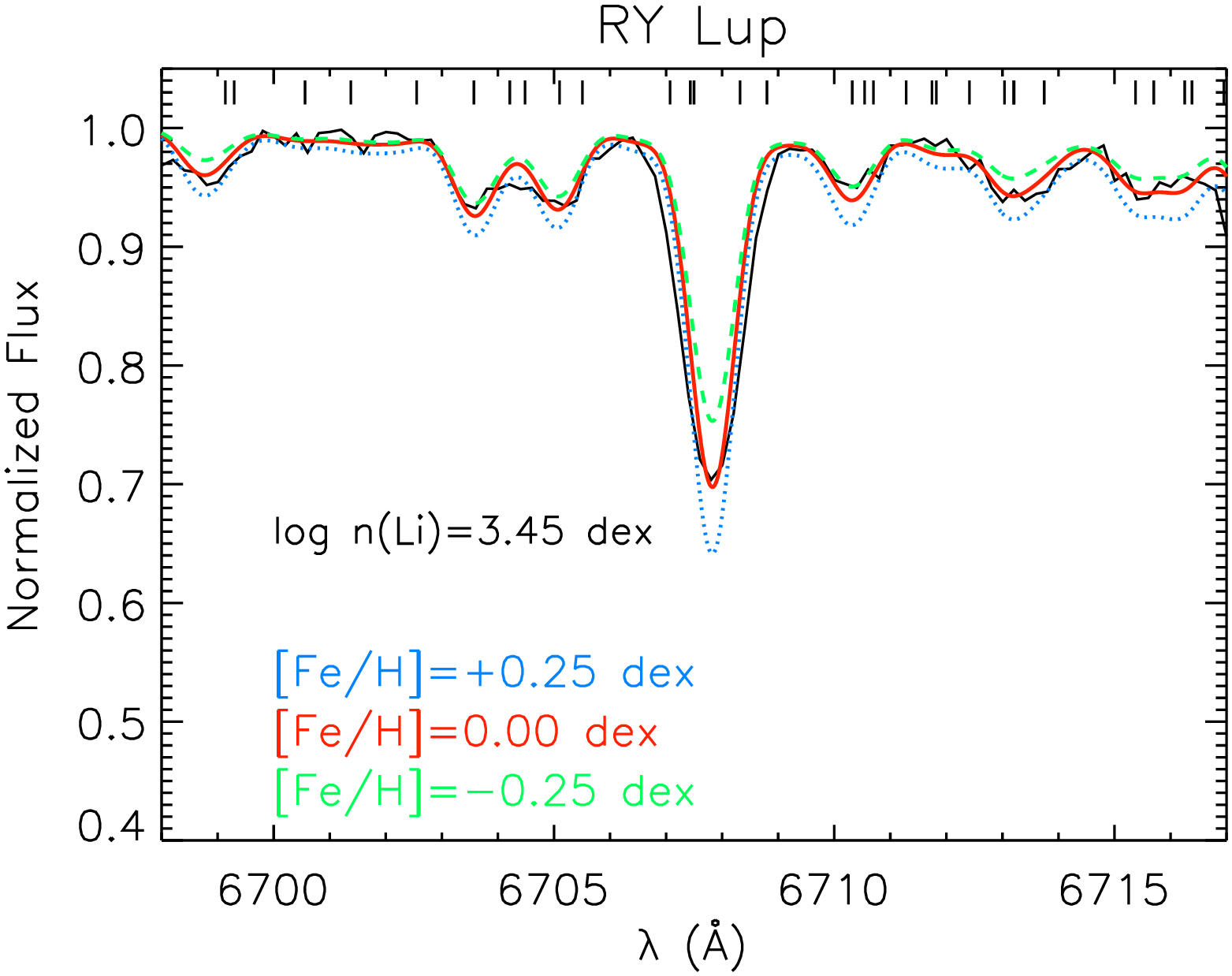}
\hspace{.5cm}
\includegraphics[width=5.5cm]{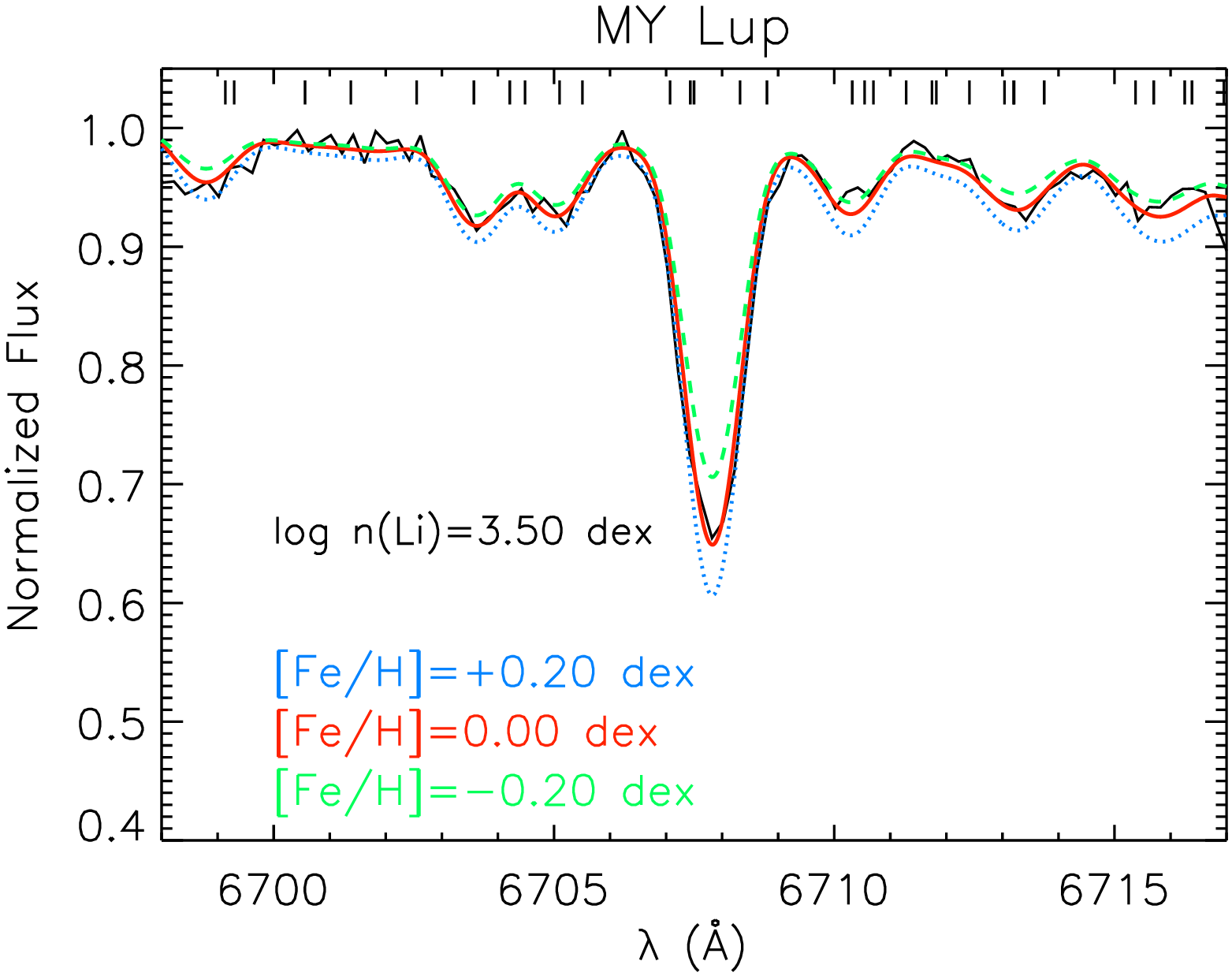}
\\
\vspace{.5cm}
\includegraphics[width=5.5cm]{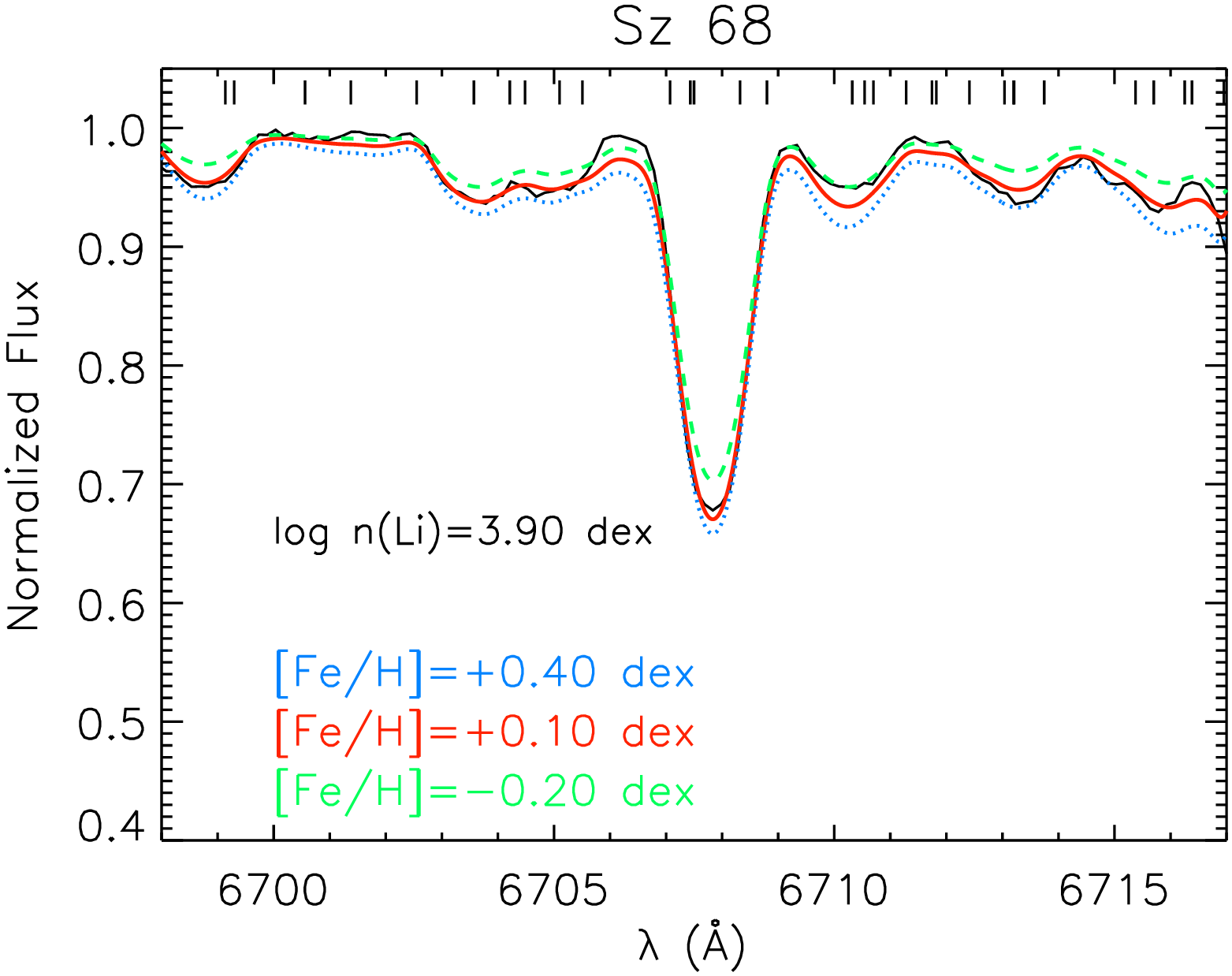}
\hspace{.5cm}
\includegraphics[width=5.5cm]{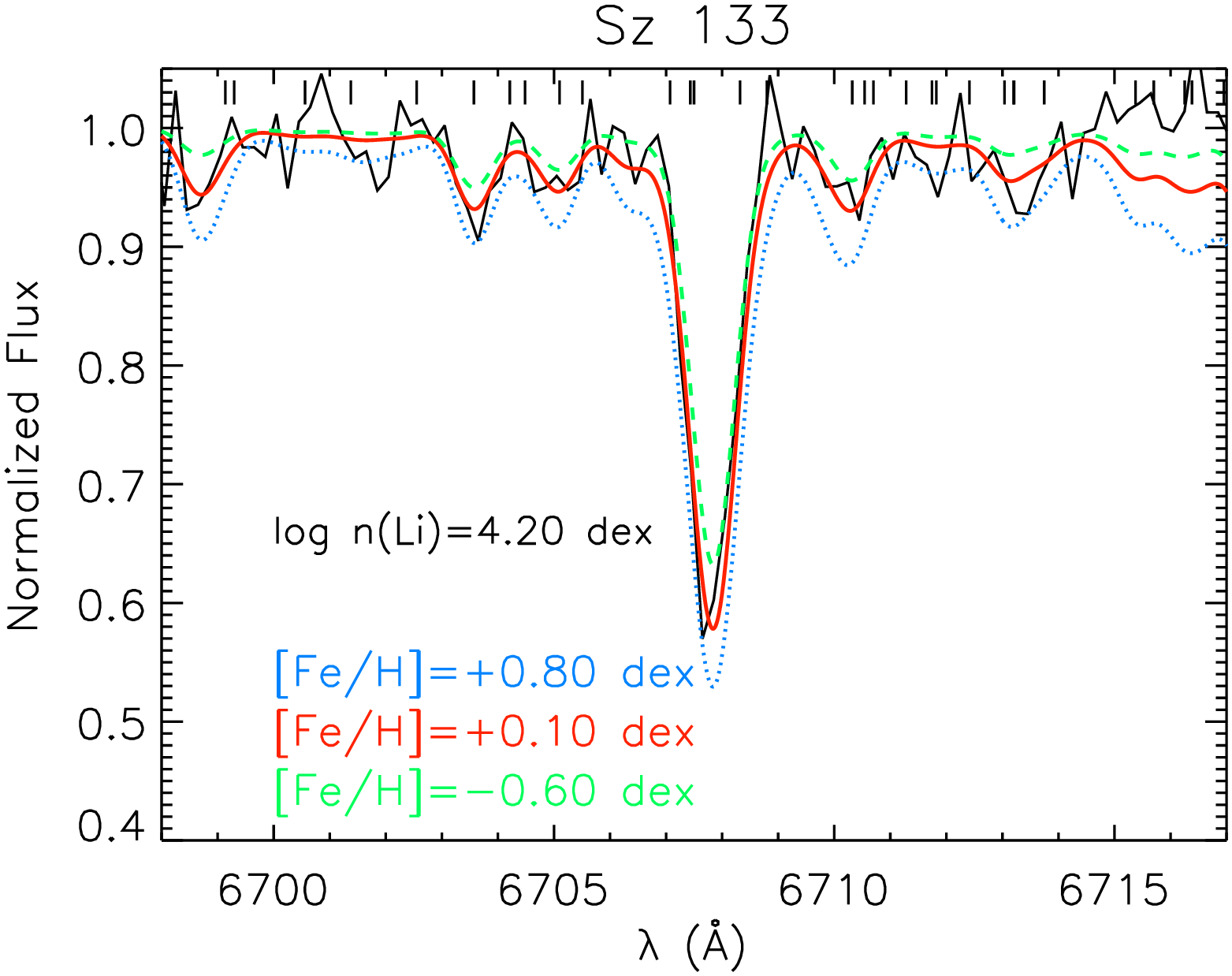}
\hspace{.5cm}
\includegraphics[width=5.5cm]{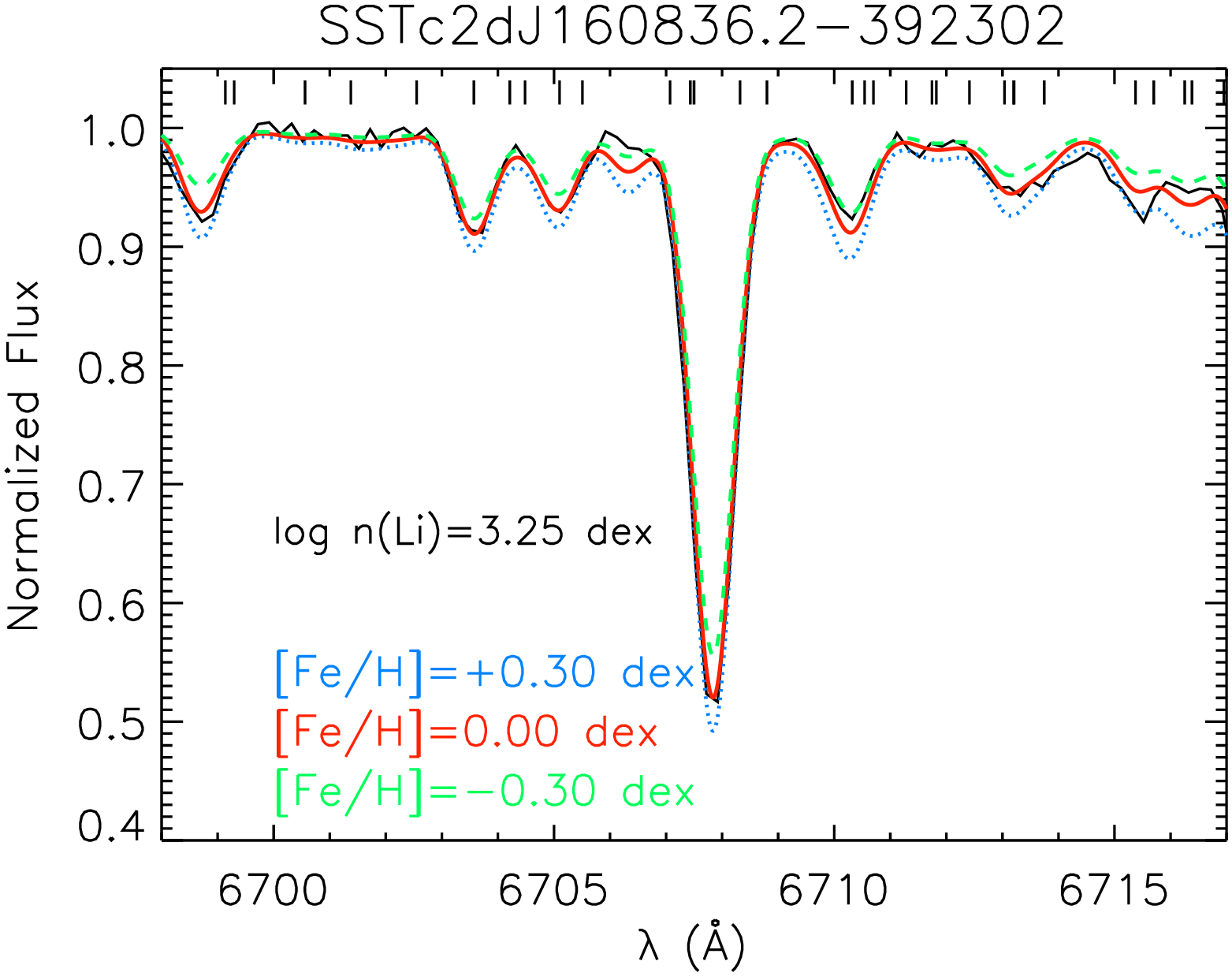}
\caption{Observed spectrum (black line) and best-fit synthetic spectrum (red line) for the six targets analyzed for abundance measurements. The blue dashed 
and green dotted lines represent the [Fe/H] values due to uncertainties in the best-fit procedure (see text). 
Iron lines are indicated with short vertical lines in the upper part of each panel.}
\label{fig:Li_spectral_synthesis} 
\end{center}
\end{figure*}

For the six selected targets, we fixed within the MOOG code the appropriate spectral resolution, the limb-darkening coefficients (taken from 
\citealt{claretetal2012}), the veiling $r_{6700}$, and the microturbulence $\xi=1.5$\,km/s, which is typical of young stars similar to 
our targets (see, e.g., \citealt{dorazietal2011, biazzoetal2011a, biazzoetal2011b}). Then, we progressively changed 
the iron abundance until the best-fit (minimum of residuals) to the observed spectrum was obtained (see Fig.\,\ref{fig:Li_spectral_synthesis}). 
As a by-product, the spectral synthesis around the Li line also allowed us to estimate the Li abundance, $A({\rm Li})^{\rm synth}$, for the six targets, 
that is found to be consistent within errors with the one obtained from the equivalent widths and the COGs (see Table\,\ref{tab:elemental_abundances} 
and Fig.\,\ref{fig:Li_spectral_synthesis}). This allowed us to make an independent assessment of the lithium abundance for these stars. 

The Fe abundance uncertainties are related both to the uncertainty in the best-fit model (we call this 
$\sigma_1$) and the errors in stellar parameters (which we call $\sigma_2$). Besides 
the uncertainty coming from the best-fit, $\sigma_1$ also includes errors in the continuum placement, that we estimated to be twice as 
large as the standard deviation obtained for the fit. We found that, at the average $T_{\rm eff}$ of 4800 K of our targets, 
$\sigma_2$ varies between 0.07 dex and 0.11 dex for typical uncertainties of $\pm$120~K 
and $\pm$0.2~dex in $T_{\rm eff}$ and $\log g$, respectively. Microturbulence velocity was fixed at 1.5 km/s, but ty\-pi\-cal 
errors of $\sim$0.2 km/s lead to a [Fe/H] uncertainty of $\pm 0.02$ dex. Other sources of error are the 
indetermination in $v \sin i$ and $r_{6700}$, which were fixed in our analysis. An uncertainty in $v \sin i$ of about 3 km/s 
may lead to errors of $\pm 0.06$ dex in iron abundance. The last source of uncertainty is veiling, that could be the largest 
for highly veiled spectra. However, for the low veiling values of the six targets here analyzed, an uncertainty of 
$\sim$0.1 in $r_{6700}$ translates into an error in [Fe/H] of $\pm 0.05$ dex. 
Final errors can be obtained by summing in quadrature the uncertainties from spectral 
synthesis and stellar parameters, plus additional contributions from rotational velocity and veiling. Typical uncertainties of $0.2-0.7$\,dex for 
[Fe/H] are derived, depending on target. The major source of error is the best-fit procedure, mainly because of the uncertainties due to 
continuum placement.

Systematic (external) errors, caused for instance by the code and/or model atmosphere, should not strongly influence our abundance analysis, as 
widely discussed by \cite{dorazietal2011}.

\subsubsection{Barium abundance}
\label{sec:barium_abun}

We determined the abundance of the $s$-process element ba\-rium through spectral synthesis of the \ion{Ba}{ii} line at 
$\lambda=5853.7$\,\AA. We checked for other Ba lines, such as the \ion{Ba}{ii} lines at 
$\lambda$4554\,\AA, $\lambda$6141\,\AA, $\lambda$6496\,\AA, and the \ion{Ba}{i} line at $\lambda$5535\,\AA, but all of them are affected by significant 
NLTE effects. Moreover, the second \ion{Ba}{ii} line and the \ion{Ba}{i} line are also strongly blended with iron lines (\citealt{mashonkinagehren2000, 
reddylambert2015, korotinetal2015}). We also tried to analyze other suitable lines of additional $s$-process elements (e.g., \ion{Y}{ii} $\lambda\lambda$4398, 
5087, 5200, 5205 \AA, \ion{Ce}{ii} $\lambda\lambda$4073, 4350, 4562 \AA, \ion{Zr}{ii} $\lambda\lambda$4161, 4209, 4443 \AA, \ion{La}{ii} $\lambda\lambda$4087, 
4322, 4333 \AA), but unfortunately with unsuccessful results, the reasons being: $i)$ the lines are too weak and/or blended with other nearby lines (due 
to the relatively low resolution of our data and the stellar rotation); $ii)$ most of those lines are in the UV-Blue (UVB) arm of the X-shooter spectrograph, 
where the continuum placement has a strong impact on the abundance estimate; $iii)$ the veiling contribution, even if low, and added to the aforementioned 
effects, makes the abundance measurement very uncertain. In addition, the $S/N$ ratio of the spectra is lower in the UVB arm than in the other spectrograph arms

Although the Ba line at $\lambda5853.7$\,\AA\,does not experience severe hyperfine structure (hfs) and isotopic 
shifts, we decided to include them and employ those provided by \cite{mcwilliam1998} with the aim of obtaining the best possible result with the spectra at 
our disposal. We then adopted the isotopic solar mixtures by \cite{andersgrevesse1989}, that is, 2.42\% for $^{134}$Ba, 7.85\% for $^{136}$Ba, 71.94\% for 
$^{138}$Ba, 6.58\% for $^{135}$Ba, and 11.21\% for $^{137}$Ba. As expected, the results do not depend on the 
abundance ratio for even and odd isotopes. As for the iron abundance, we considered the solar barium abundance by \cite{asplundetal2009}, that is,
$\log n({\rm Ba})_\odot = 2.18 \pm 0.09$\,dex, and the stellar parameters by \cite{frascaetal2017}. Then, we ran MOOG fixing the 
spectral resolution, the limb-darkening coefficients (taken by \citealt{claretetal2012}), the veiling $r_{5800}$ (assuming a mean 
value between $r_{5200}$ and $r_{6200}$; see Table\,\ref{tab:param}), and the microturbulence $\xi=1.5$\,km/s. The fitting procedure and the error 
estimates are the same as for the iron abundance determination described in the previous Section, and for the same six selected stars.

\begin{figure*}[t!]
\begin{center}
\includegraphics[width=5cm]{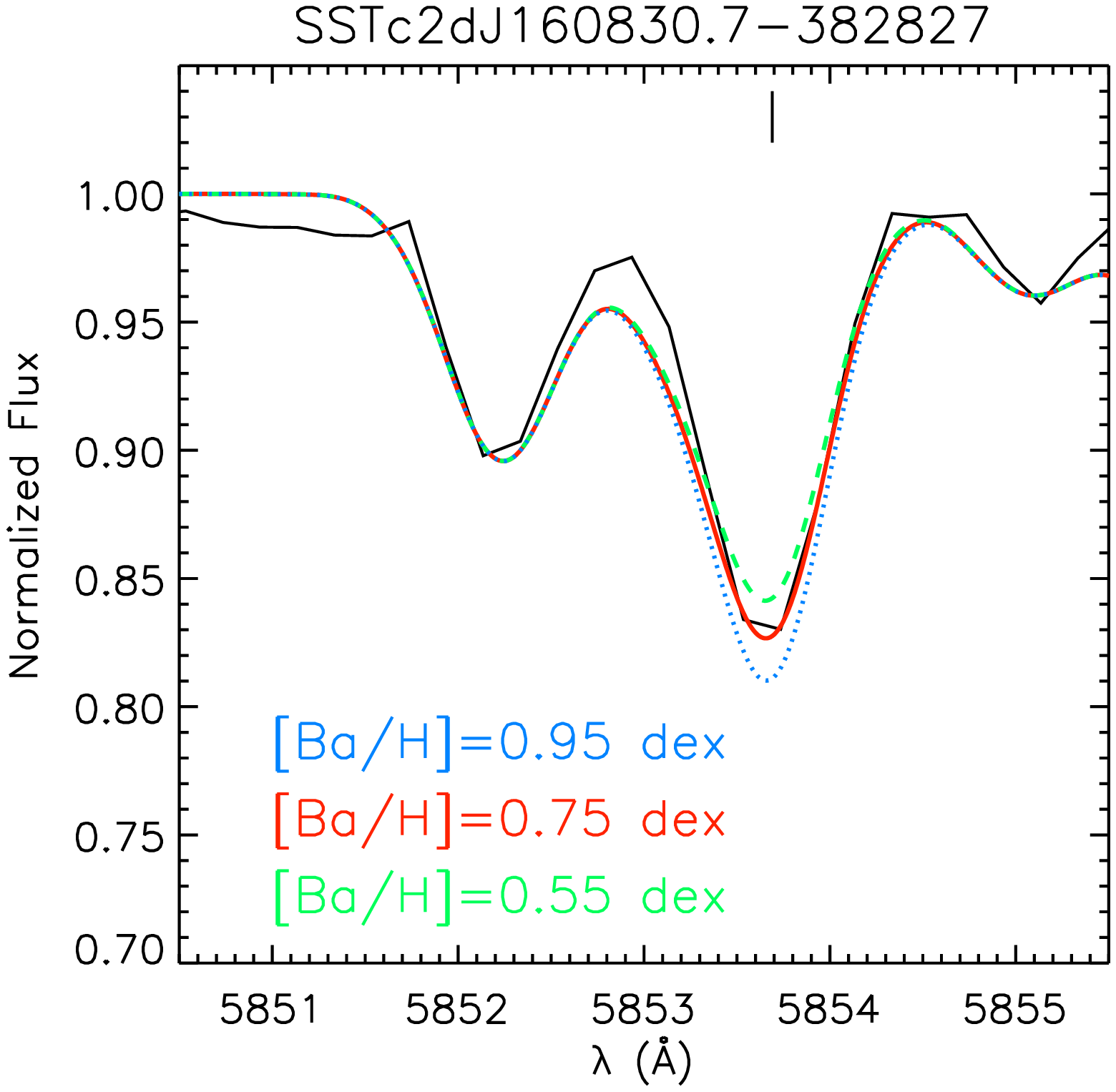}
\hspace{.5cm}
\includegraphics[width=5cm]{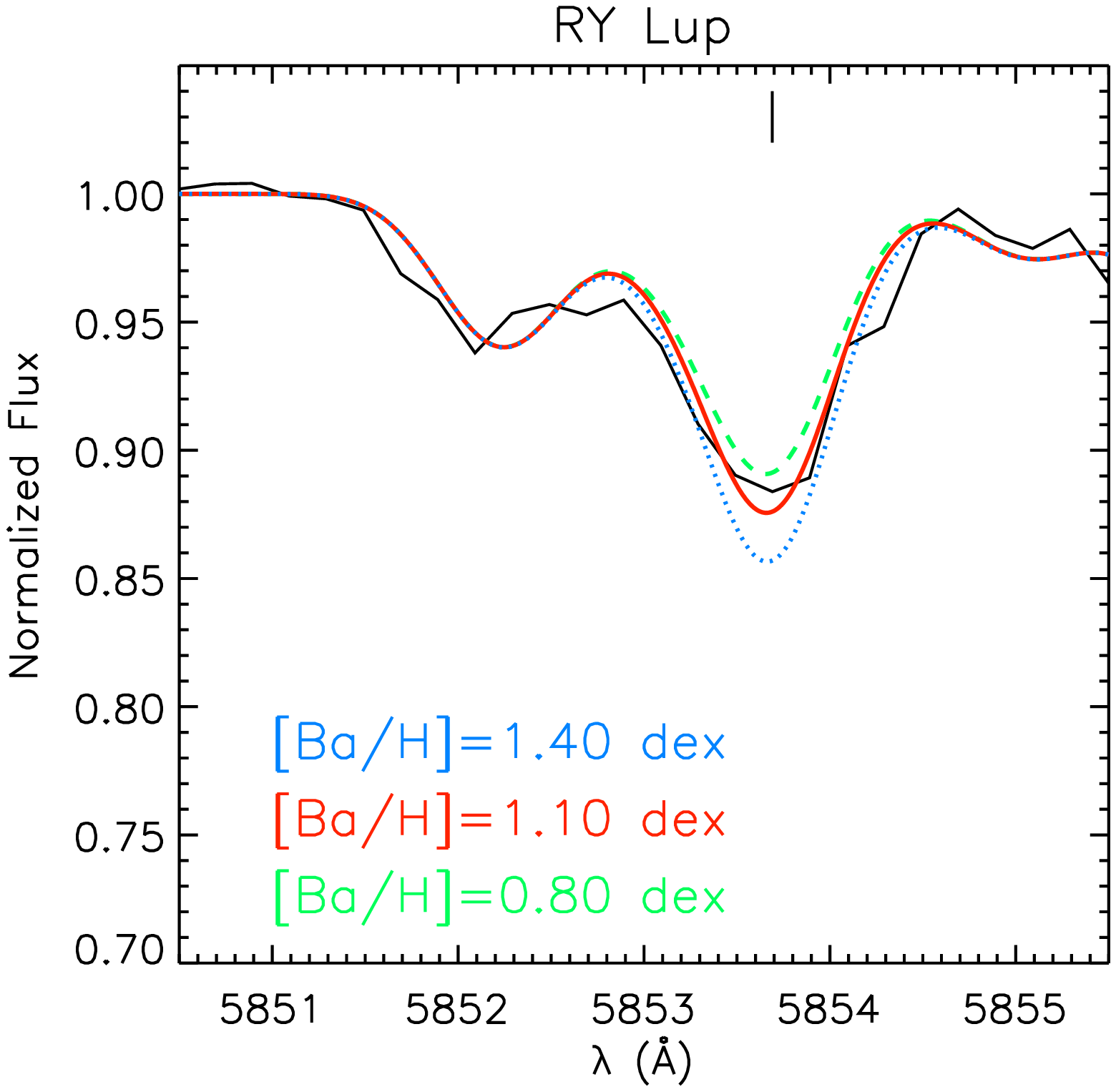}
\hspace{.5cm}
\includegraphics[width=5cm]{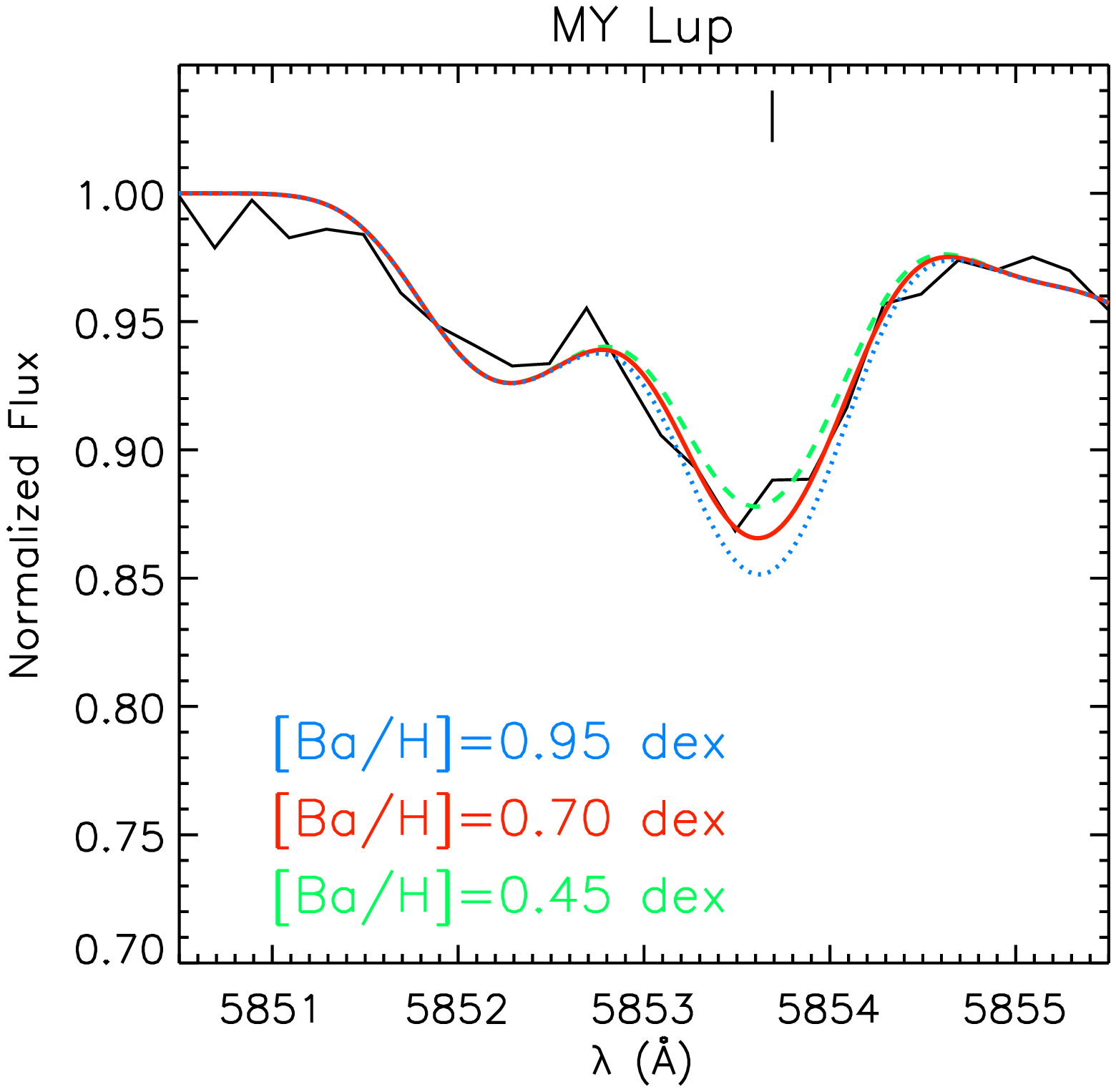}
\\
\vspace{.5cm}
\includegraphics[width=5cm]{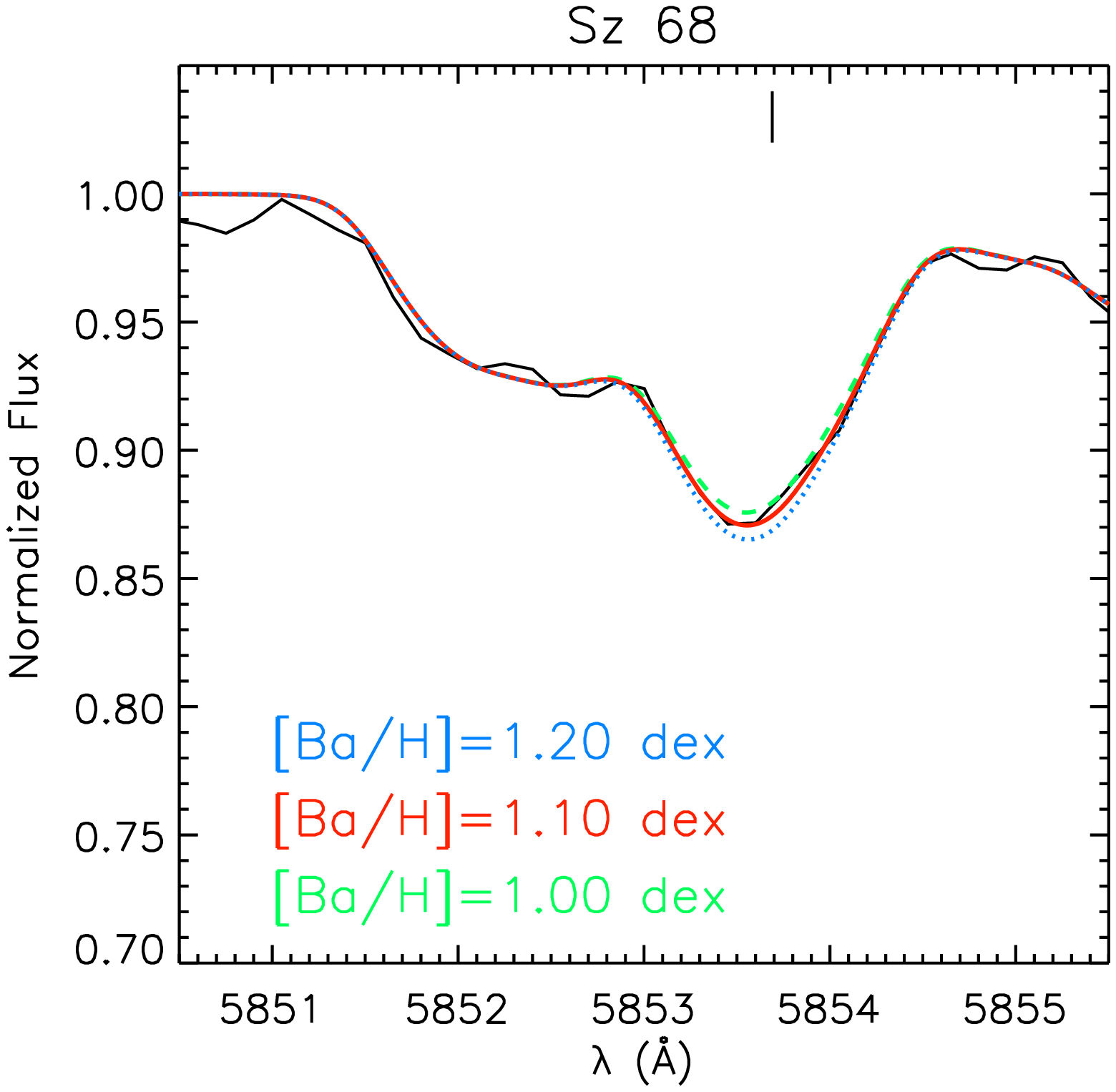}
\hspace{.5cm}
\includegraphics[width=5cm]{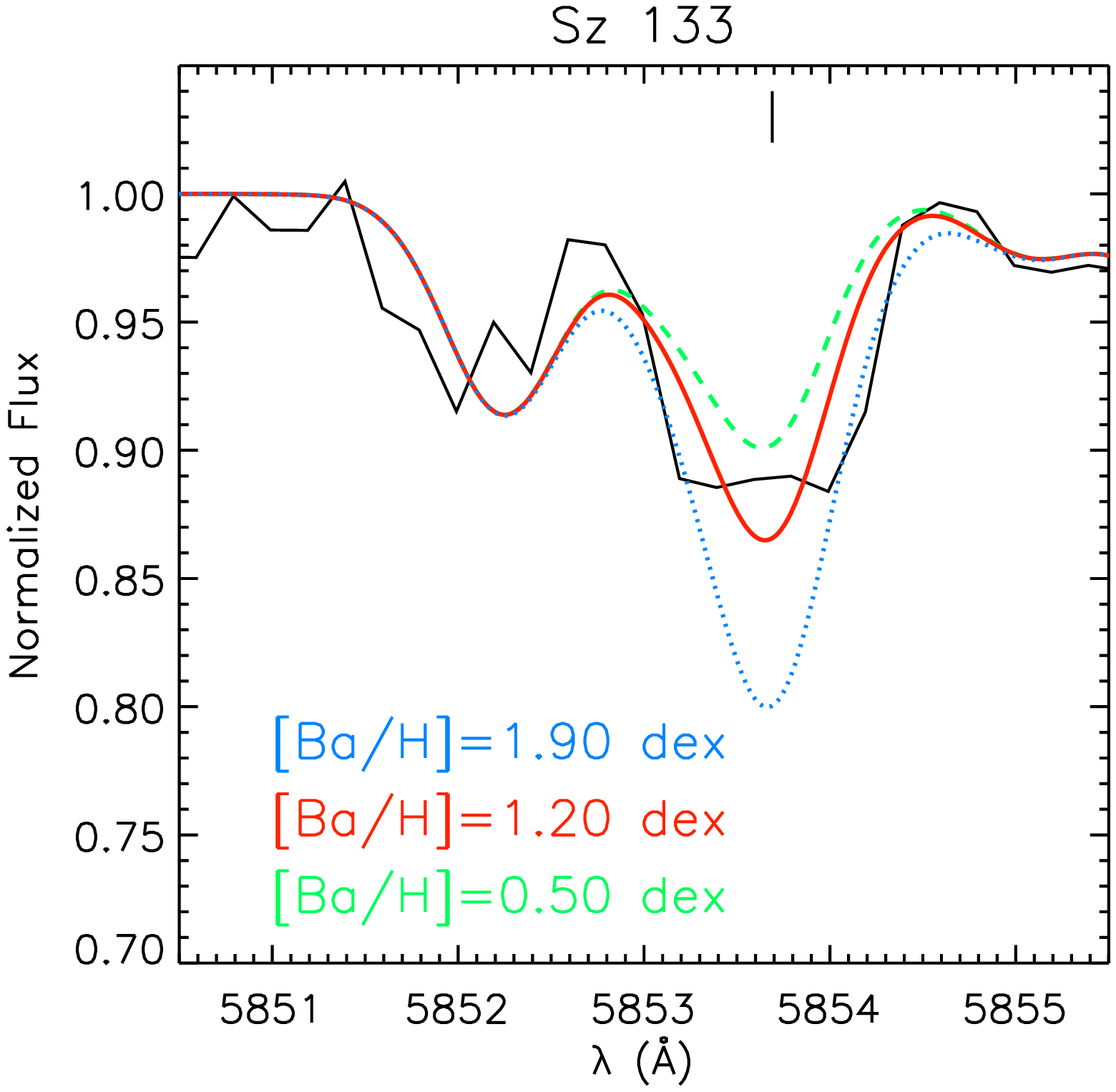}
\hspace{.5cm}
\includegraphics[width=5cm]{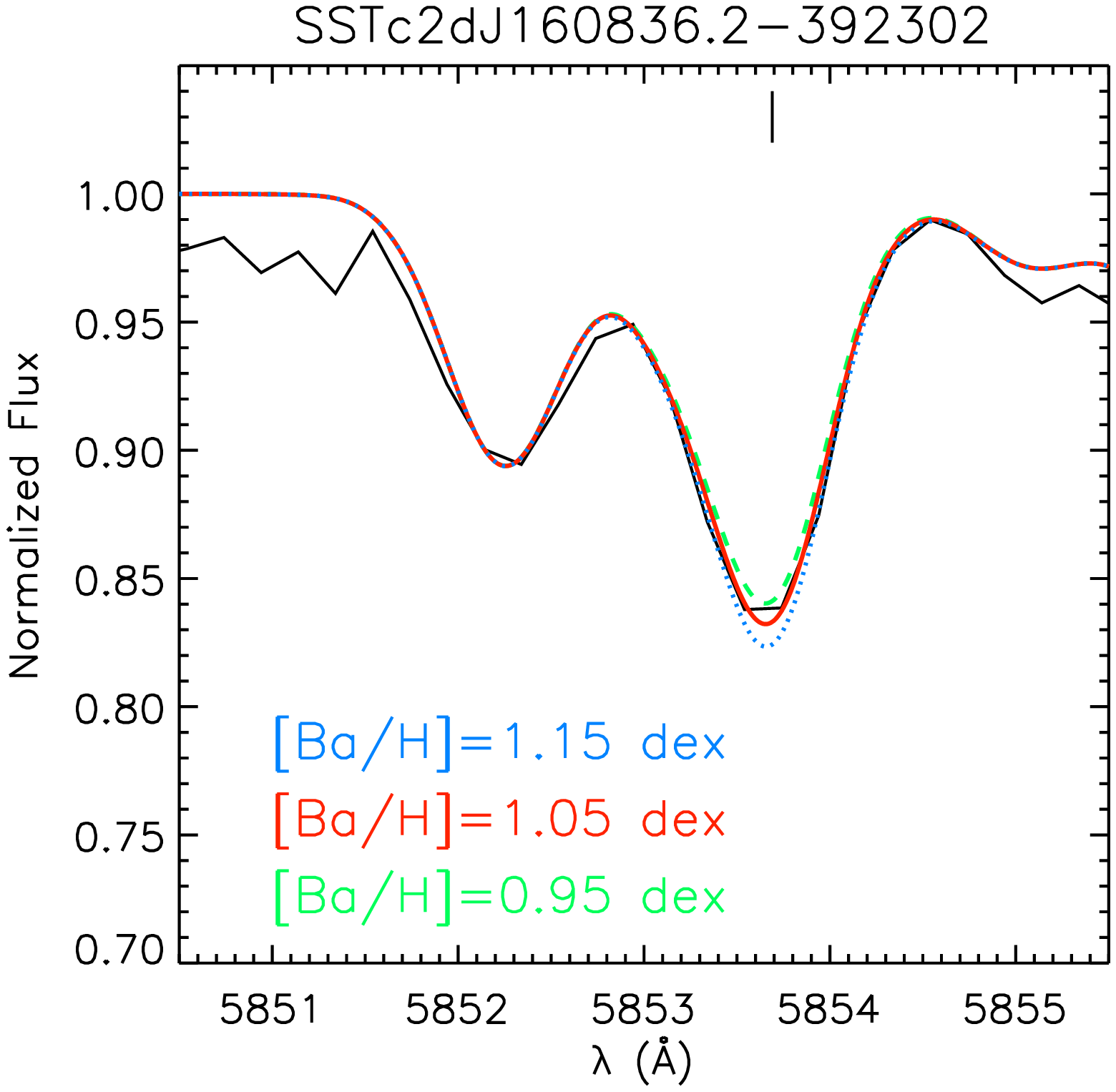}
\caption{Visualization of spectral synthesis as in Fig.\,\ref{fig:Li_spectral_synthesis} but for the \ion{Ba}{ii} line at $\lambda5853.7$\,\AA. 
The barium line is marked with a short vertical line in the upper part of each panel.}
\label{fig:Ba_spectral_synthesis} 
\end{center}
\end{figure*}

In Fig.\,\ref{fig:Ba_spectral_synthesis} we show the spectral synthesis of the six targets in the wavelength region around the \ion{Ba}{ii} line, 
with the corresponding error $\sigma_1$ (see Table\,\ref{tab:elemental_abundances}). As done for the Fe abundance, to evaluate the impact of the variation 
of the stellar parameters ($T_{\rm eff}$, $\log g$, and $\xi$) and of $v \sin i$ and $r$, we varied each quantity separately (leaving the others unchanged) 
and checked the abundance sensitivity to that variation. A change of $\pm$120 K in $T_{\rm eff}$, $\pm$0.2 dex in $\log g$, and $\pm$0.2 km/s in $\xi$, leads 
to Ba abundance variations of 0.07 dex, 0.03 dex, and 0.15 dex, respectively. This means that the barium abundance measurement is strongly influenced by the 
microturbulence (also noted by other authors, for example, \citealt{dorazietal2012}), as the barium line is close to the flat part of the curve of growth. 
Variations of 3 km/s in $v \sin i$ lead to 0.01 dex uncertainty in [Ba/H], while an uncertainty of about 0.1 in veiling yields an error of 
$\sim$0.15\,dex in [Ba/H]. Therefore, veiling also has a strong influence on this element. We will discuss the impact of $\xi$ and $r_{5800}$ measurement 
on the determination of barium abundance in Sect.\,\ref{sec:barium_issue}. 
The cumulative uncertainty in [Ba/H] can be obtained by summing in quadrature the uncertainties from the fit, those 
on the stellar parameters, and the additional contribution of the $v \sin i$ error (negligible) and veiling. The 
cumulative uncertainties in [Ba/H] measurements for our targets are of $\sim$0.2-0.7\,dex, that is, they are dominated by the best-fit procedure. 

\setlength{\tabcolsep}{3pt}

\begin{table*}[htb]
\caption[]{Stellar parameters. Object name, spectral type, effective temperature, surface gravity, rotational velocity, radial velocity, veiling at 
$\lambda$=5400~\AA, $6200$~\AA, and 7100~\AA, measured and corrected lithium equivalent width and lithium abundance.} 
\scriptsize
\begin{tabular}{llllrrrrrrrr}
\hline
\hline
\noalign{\smallskip}
 Name	                  & SpT & $T_{\rm eff}$ &   $\log g$   &  $v \sin i$  & $V_{\rm rad}$ & $r_{5400}$& $r_{6200}$& $r_{7100}$&$EW_{\rm Li}$ &$EW_{\rm Li}^{\rm corr}$&$A$(Li)  \\
                          &                            &    (K)       &    (dex)      &    (km/s)    &    (km/s)     &	     &  	 &	     & (m\AA)	   & (m\AA) & (dex) \\
\hline
\noalign{\smallskip}
\multicolumn{12}{c}{Sample of class II YSOs by \cite{alcalaetal2014}}\\
\hline
  Sz\,66  		  & M3   &3351$\pm$ 47  & 3.81$\pm$0.21 &  $\leq$8.0   & 17.4$\pm$1.8  &    1.5 &	  0.8 & $\leq$0.2 &  370$\pm$10 &481  &2.86$\pm$0.12  \\ 
  AKC2006\,19		  & M5   &3027$\pm$ 34  & 4.45$\pm$0.11 &  $\leq$8.0   &  9.6$\pm$2.1  &    ... & $\leq$0.2 & $\leq$0.2 &  660$\pm$21 &631  &3.34$\pm$0.04  \\ 
  Sz\,69  		  & M4.5 &3163$\pm$119  & 3.50$\pm$0.23 & 33.0$\pm$4.0 &  5.4$\pm$2.9  &    ... & $\leq$0.2 & $\leq$0.2 &  150$\pm$10 &122  &     $< -1$    \\
  Sz\,71  		  & M1.5 &3599$\pm$ 35  & 4.27$\pm$0.24 &  $\leq$8.0   & 11.7$\pm$1.9  &    0.6 &	  0.5 &       0.3 &  540$\pm$12 &720  &3.00$\pm$0.16  \\ 
  Sz\,72  		  & M2   &3550$\pm$ 70  & 4.18$\pm$0.28 &  $\leq$8.0   &  6.9$\pm$2.4  &    ... & $\leq$0.2 & $\leq$0.2 &  418$\pm$10 &392  &1.87$\pm$0.21  \\ 
  Sz\,73  		  & K7   &3980$\pm$ 33  & 4.40$\pm$0.33 & 31.0$\pm$3.0 &  5.0$\pm$2.2  &    1.0 &	  1.0 &       0.5 &  527$\pm$11 &832  &3.32$\pm$0.08   \\
  Sz\,74  		  & M3.5 &3371$\pm$ 79  & 3.98$\pm$0.11 & 30.2$\pm$1.0 &  1.0$\pm$1.5  &    0.7 &	   0.4 & $\leq$0.2 &  472$\pm$12 &535  &3.39$\pm$0.20	\\
  Sz\,83  		  & K7   &4037$\pm$ 96  & 3.5 $\pm$0.4  &  8.5$\pm$4.8 &  3.3$\pm$1.8  &    ... &	   3.5 &       2.4 &  272$\pm$10 &988  &  $>$3.81     \\ 
  Sz\,84$^a$  		  & M5   &3058$\pm$ 82  & 4.20$\pm$0.21 & 21.3$\pm$4.1 & 11.9$\pm$2.0  &    0.9 &	   0.3 & $\leq$0.2 &  490$\pm$12 &531  &2.69$\pm$0.16  \\
  Sz\,130 		  & M2   &3448$\pm$ 62  & 4.48$\pm$0.10 &  $\leq$8.0   &  3.6$\pm$2.6  &    ... &	   0.7 & $\leq$0.2 &  455$\pm$10 &579  &     $>$3.5   \\ 
  Sz88\,A		  & M0   &3700$\pm$ 19  & 3.77$\pm$0.47 &  $\leq$8.0   &  6.5$\pm$2.3  &    ... &	   2.1 &       1.4 &  308$\pm$10 &777  &3.17$\pm$0.23  \\
  Sz88\,B		  & M4.5 &3075$\pm$111  & 4.08$\pm$0.12 &  $\leq$8.0   &  5.7$\pm$2.1  &    ... &	   0.3 &       0.3 &  570$\pm$10 &704  &  $>$3.50     \\ 
  Sz\,91$^a$  		  & M1   &3664$\pm$ 45  & 4.34$\pm$0.22 &  $\leq$8.0   & 19.0$\pm$2.4  &    0.9 &	   0.6 &       0.4 &  480$\pm$10 &681  &3.00$\pm$0.22  \\
  Lup\,713		  & M5.5 &2943$\pm$109  & 3.89$\pm$0.13 & 24.0$\pm$4.0 &  3.9$\pm$3.3  &    ... &	   ... & $\leq$0.2 &  352$\pm$25 &323  &1.04$\pm$0.20  \\ 
  Lup\,604s		  & M5.5 &3014$\pm$ 62  & 3.89$\pm$0.12 & 31.5$\pm$2.8 &  2.7$\pm$1.9  &    ... & $\leq$0.2 & $\leq$0.2 &  575$\pm$25 &546  &2.71$\pm$0.21  \\ 
  Sz\,97  		  & M4   &3185$\pm$ 78  & 4.14$\pm$0.16 & 25.1$\pm$1.5 &  2.4$\pm$2.2  &    0.7 & $\leq$0.2 & $\leq$0.2 &  525$\pm$15 &497  &2.64$\pm$0.20	\\
  Sz\,99  		  & M4   &3297$\pm$ 98  & 3.89$\pm$0.21 & 38.0$\pm$3.0 &  3.2$\pm$3.1  &    ... & $\leq$0.2 & $\leq$0.2 &  213$\pm$15 &186  &0.43$\pm$0.14  \\ 
  Sz\,100$^a$ 		  & M5.5 &3037$\pm$ 44  & 3.87$\pm$0.21 & 16.6$\pm$5.4 &  2.7$\pm$2.5  &    ... & $\leq$0.2 & $\leq$0.2 &  525$\pm$15 &496  &2.39$\pm$0.14	\\
  Sz\,103 		  & M4   &3380$\pm$ 36  & 3.97$\pm$0.10 & 12.0$\pm$4.0 &  1.4$\pm$2.2  &    0.7 &	   0.3 & $\leq$0.2 &  517$\pm$15 &564  &     $>$3.5   \\
  Sz\,104 		  & M5   &3074$\pm$ 73  & 3.96$\pm$0.10 &  $\leq$8.0   &  2.3$\pm$2.3  &    ... & $\leq$0.2 & $\leq$0.2 &  570$\pm$24 &542  &2.83$\pm$0.24  \\ 
  Lup\,706$^b$		  & M7.5 &2750$\pm$ 82  & 3.93$\pm$0.11 & 25.0$\pm$15.0& 11.9$\pm$4.5  &    ... &	   ... & $\leq$0.2 &   $<$100$^d$&$<$69&     $< -1$   \\
  Sz\,106$^b$ 		  & M0.5 &3691$\pm$ 35  & 4.82$\pm$0.13 &  $\leq$8.0   &  8.0$\pm$2.6  &    0.5 &	   0.5 & $\leq$0.2 &  515$\pm$12 &612  &  $<$2.91     \\ 
  Par-Lup3-3		  & M4   &3461$\pm$ 49  & 4.47$\pm$0.11 &  $\leq$8.0   &  3.5$\pm$2.5  &    ... & $\leq$0.2 & $\leq$0.2 &  560$\pm$25 &534  &3.20$\pm$0.11  \\ 
  Par-Lup3-4$^b$	  & M4.5 &3089$\pm$246  & 3.56$\pm$0.80 & 12.0$\pm$10.0&  5.6$\pm$4.3  &    ... &	   ... &       0.3 &   $<$100$^d$&$<$93&     $< -1$   \\
  Sz\,110 		  & M4   &3215$\pm$162  & 4.31$\pm$0.21 &  $\leq$8.0   &  2.6$\pm$2.3  &    2.7 &	   1.0 &       0.4 &  370$\pm$10 &583  &3.40$\pm$0.27  \\ 
  Sz\,111$^a$ 		  & M1   &3683$\pm$ 34  & 4.66$\pm$0.21 &  $\leq$8.0   & 13.8$\pm$2.1  &    0.6 &	   0.5 &       0.3 &  490$\pm$12 &650  &  $<$2.95     \\ 
  Sz\,112$^a$ 		  & M5   &3079$\pm$ 47  & 3.96$\pm$0.10 &  $\leq$8.0   &  6.2$\pm$1.7  &    ... & $\leq$0.2 & $\leq$0.2 &  590$\pm$12 &562  &3.03$\pm$0.12  \\
  Sz\,113 		  & M4.5 &3064$\pm$114  & 3.76$\pm$0.27 &  $\leq$8.0   &  6.1$\pm$2.4  &    ... &	   ... &       0.5 &  210$\pm$15 &272  &0.93$\pm$0.21  \\ 
 2MASS\,J16085953-3856275 & M8.5 &2649$\pm$ 31  & 3.98$\pm$0.10 &  $\leq$8.0   &  7.2$\pm$4.0  &    ... & $\leq$0.2 & $\leq$0.2 &	$<$80$^d$&$<$48&     $< -1$   \\  
SSTc2d\,160901.4-392512   & M4   &3305$\pm$ 57  & 4.51$\pm$0.11 &  $\leq$8.0   & 15.9$\pm$0.7  &    0.7 & $\leq$0.2 & $\leq$0.2 &  565$\pm$20 &538  &     $>$3.5   \\   
  Sz\,114 		  & M4.8 &3134$\pm$ 35  & 3.92$\pm$0.12 &  $\leq$8.0   &  4.0$\pm$2.4  &    0.6 &	   0.3 &       0.3 &  565$\pm$10 &698  &     $>$3.5   \\ 
  Sz\,115 		  & M4.5 &3124$\pm$ 42  & 3.90$\pm$0.21 &  9.2$\pm$6.3 &  6.4$\pm$2.3  &    ... & $\leq$0.2 & $\leq$0.2 &  579$\pm$15 &551  &0.93$\pm$0.21  \\
  Lup\,818s		  & M6   &2953$\pm$ 59  & 3.97$\pm$0.11 & 12.4$\pm$6.0 &  5.1$\pm$2.0  &    ... & $\leq$0.2 & $\leq$0.2 &  459$\pm$2  &430  &1.71$\pm$0.15	 \\
  Sz\,123\,A$^a$	  & M1   &3521$\pm$ 70  & 4.46$\pm$0.13 & 12.3$\pm$3.0 & 16.8$\pm$1.8  &    1.2 &	   0.6 & $\leq$0.2 &  410$\pm$12 &499  &2.34$\pm$0.18  \\
  Sz\,123\,B$^b$	  & M2   &3513$\pm$ 45  & 4.17$\pm$0.22 &  $\leq$8.0   &  7.4$\pm$2.3  &    0.8 &	   0.6 & $\leq$0.2 &  421$\pm$8  &513  &2.59$\pm$0.11  \\
  SST-Lup3-1		  & M5   &3042$\pm$ 43  & 3.95$\pm$0.11 &  $\leq$8.0   &  6.3$\pm$2.6  &    ... & $\leq$0.2 & $\leq$0.2 &  530$\pm$15 &501  &2.43$\pm$0.14  \\ 
~\\
\multicolumn{12}{c}{Sample of class II YSOs by \cite{alcalaetal2017}}\\
\hline
  Sz\,65                  & K7   &4005$\pm$ 75  & 3.85$\pm$0.26 &  $\leq$8.0   &$-$2.7$\pm$2.0 &	 0.3&	    0.3 & $\leq$0.2 & 602$\pm$10 &667  &3.03$\pm$0.13  \\
  AKC2006\,18 	          & M6.5 &2930$\pm$ 45  & 4.46$\pm$0.11 &  $\leq$8.0   &   9.1$\pm$2.3 &	 ...& $\leq$0.2 & $\leq$0.2 & 545$\pm$100&515  &2.23$\pm$0.72	\\ 
SSTc2d\,J154508.9-341734  & M5.5 &3242$\pm$205  & 3.33$\pm$0.77 &  $\leq$8.0   &$-$0.8$\pm$2.7 &	 ...&	    0.8 &	0.7 &	$<$60$^d$&$<$57&     $< -1$    \\
  Sz\,68                  & K2   &4506$\pm$ 82  & 3.68$\pm$0.12 & 39.6$\pm$1.2 &$-$4.3$\pm$1.8 &	 0.3&	    0.3 &	0.4 & 441$\pm$10 &571  &3.64$\pm$0.12	 \\
SSTc2d\,J154518.5-342125  & M6.5 &2700$\pm$100  & 3.45$\pm$0.11 & 13.0$\pm$6.0 &   4.4$\pm$2.9 &	 ...& $\leq$0.2 & $\leq$0.2 & 525$\pm$15 &493  &    $<$1.87   \\ 
  Sz81\,A                 & M4.5 &3077$\pm$151  & 3.48$\pm$0.11 &  $\leq$8.0   &$-$0.1$\pm$2.9 &	 ...&	    0.3 & $\leq$0.2 & 510$\pm$20 &554  &2.96$\pm$0.25	\\ 
  Sz81\,B                 & M5.5 &2991$\pm$ 76  & 3.53$\pm$0.11 & 25.0$\pm$5.0 &   1.2$\pm$2.4 &	 ...&	    0.3 & $\leq$0.2 & 445$\pm$15 &478  &2.14$\pm$0.17	 \\
  Sz\,129                 & K7   &4005$\pm$ 45  & 4.49$\pm$0.25 &  $\leq$10.0  &   3.2$\pm$2.5 &	 1.4&	    1.0 & $\leq$0.2 & 440$\pm$11 &627  &2.76$\pm$0.05  \\ 
SSTc2d\,J155925.2-423507  & M5   &2984$\pm$ 82  & 4.41$\pm$0.12 &  $\leq$8.0   &   6.5$\pm$2.5 &	 ...& $\leq$0.2 & $\leq$0.2 & 660$\pm$40 &631  &3.37$\pm$0.05  \\ 
  RY\,Lup$^a$             & K2   &5082$\pm$118  & 3.87$\pm$0.22 & 16.3$\pm$5.3 &   1.3$\pm$2.0 &	 0.4&	    0.5 & $\leq$0.2 & 377$\pm$10 &455  &3.70$\pm$0.12	\\
SSTc2d\,J160000.6-422158  & M4.5 &3086$\pm$ 82  & 4.03$\pm$0.11 &  $\leq$8.0   &   2.5$\pm$2.2 &	 ...& $\leq$0.2 & $\leq$0.2 & 580$\pm$20 &552  &2.96$\pm$0.15	\\ 
SSTc2d\,J160002.4-422216  & M4   &3159$\pm$140  & 4.00$\pm$0.48 &  $\leq$8.0   &   2.6$\pm$2.6 &	 0.3& $\leq$0.2 & $\leq$0.2 & 660$\pm$80 &632  &   $>$3.5   \\ 
SSTc2d\,J160026.1-415356  & M5.5 &2976$\pm$221  & 3.97$\pm$0.35 &  $\leq$8.0   &$-$1.1$\pm$2.3 &	 ...&	    0.3 & $\leq$0.2 & 560$\pm$50 &610  &3.22$\pm$0.58  \\ 
  MY\,Lup$^a$             & K0   &4968$\pm$200  & 3.72$\pm$0.18 & 29.1$\pm$2.0 &   4.4$\pm$2.1 &$\leq$0.2& $\leq$0.2 & $\leq$0.2 & 468$\pm$10 &454  &3.67$\pm$0.17  \\
  Sz\,131	          & M3   &3300$\pm$122  & 4.29$\pm$0.36 &  $\leq$8.0   &   2.4$\pm$2.2 &	 ...&	    0.3 & $\leq$0.2 & 535$\pm$30 &584  &     $>$3.5   \\ 
  Sz\,133$^b$             & K5   &4420$\pm$129  & 3.96$\pm$0.51 &  $\leq$8.0   &   0.7$\pm$2.5 &	 0.7&	    0.7 &	0.5 & 420$\pm$25 &643  &3.56$\pm$0.20  \\ 
SSTc2d\,J160703.9-391112$^b$&M4.5&3072$\pm$ 55 & 4.01$\pm$0.11 &  $\leq$8.0   &   1.8$\pm$2.7 &     ... & $\leq$0.2 & $\leq$0.2 & 680$\pm$50 &652  &  $>$3.50     \\   
SSTc2d\,J160708.6-391408$^c$&... &3474$\pm$206 & 4.18$\pm$0.56 &  $\leq$8.0   &$-$4.2$\pm$2.9 &     ... &	   1.0 &       0.9 & 504$\pm$65 &932  &  $>$3.50     \\	
  Sz\,90                  & K7   &4022$\pm$ 52  & 4.24$\pm$0.42 &  $\leq$8.0   &   1.6$\pm$2.3 &	0.5&	   0.5 & $\leq$0.2 & 491$\pm$10 &587  &2.71$\pm$0.08  \\ 
  Sz\,95                  & M3   &3443$\pm$ 53  & 4.37$\pm$0.12 &  $\leq$8.0   &$-$2.8$\pm$2.3 &	...& $\leq$0.2 & $\leq$0.2 & 585$\pm$10 &559  &3.49$\pm$0.02  \\ 
  Sz\,96                  & M1   &3702$\pm$ 57  & 4.50$\pm$0.23 &  $\leq$8.0   &$-$2.7$\pm$2.6 &	0.3&	   0.4 & $\leq$0.2 & 595$\pm$10 &683  &2.99$\pm$0.13  \\ 
2MASS\,J16081497-3857145$^a$&M5.5&3024$\pm$ 46 & 3.96$\pm$0.23 & 22.0$\pm$4.2 &   6.0$\pm$2.9 &	...& $\leq$0.2 & $\leq$0.2 & 525$\pm$100&496  &2.43$\pm$0.74   \\ 
  Sz\,98                  & K7   &4080$\pm$ 71  & 4.10$\pm$0.23 &  $\leq$8.0   &$-$1.4$\pm$2.1 &	 0.5&	    0.6 & $\leq$0.2 & 508$\pm$10 &633  &2.97$\pm$0.11  \\ 
  Lup\,607                & M6.5 &3084$\pm$ 46  & 4.48$\pm$0.11 &  $\leq$8.0   &   6.8$\pm$2.4 &	 ...& $\leq$0.2 & $\leq$0.2 & 539$\pm$20 &511  &2.43$\pm$0.11	\\ 
  Sz\,102$^b$ 	          & K2   &5145$\pm$ 50  & 4.10$\pm$0.50 & 41.0$\pm$2.8 &  21.6$\pm$12.4&	 ...&	    2.5 &	2.0 & 196$\pm$50 &595  &3.98$\pm$0.18	\\  
SSTc2d\,J160830.7-382827  & K2   &4797$\pm$145  & 4.09$\pm$0.23 &  $\leq$8.0   &   1.2$\pm$1.9 &	 0.1& $\leq$0.2 & $\leq$0.2 & 436$\pm$10 &421  &3.40$\pm$0.20	\\ 
SSTc2d\,J160836.2-392302$^a$&K6  &4429$\pm$ 83 & 4.04$\pm$0.13 &  $\leq$8.0   &   2.2$\pm$2.1 &	0.3&	   0.3 & $\leq$0.2 & 454$\pm$35 &501  &3.09$\pm$0.18 \\   
  Sz\,108\,B              & M5   &3102$\pm$ 59  & 3.86$\pm$0.22 &  $\leq$8.0   &   0.2$\pm$2.2 &	 ...& $\leq$0.2 & $\leq$0.2 & 495$\pm$25 &467  &2.28$\pm$0.16  \\ 
2MASS\,J16085324-3914401  & M3   &3393$\pm$ 85  & 4.24$\pm$0.41 &  $\leq$8.0   &   0.9$\pm$2.2 &	 ...&	    0.3 & $\leq$0.2 & 577$\pm$12 &663  &3.50$\pm$0.03  \\ 
2MASS\,J16085373-3914367  & M5.5 &2840$\pm$200  & 3.60$\pm$0.45 &  $\leq$8.0   &   6.1$\pm$6.9 &	 ...& $\leq$0.2 & $\leq$0.2 &$<$100$^d$&$<$70  &   $< -1$      \\ 
2MASS\,J16085529-3848481  & M4.5 &2899$\pm$ 55  & 3.99$\pm$0.11 &  $\leq$8.0   &$-$1.1$\pm$2.4 &	 ...& $\leq$0.2 & $\leq$0.2 & 591$\pm$40 &561  &2.51$\pm$0.41  \\ 
SSTc2d\,J160927.0-383628$^a$&M4.5&3147$\pm$ 58 & 4.00$\pm$0.10 &  $\leq$8.0   &   3.4$\pm$2.2 &	...&	   1.5 &       0.5 & 132$\pm$15 &209  &     $<$0.53  \\ 
  Sz\,117                 & M3.5 &3470$\pm$ 51  & 4.10$\pm$0.32 &  $\leq$8.0   &$-$1.4$\pm$2.2 &	 0.7&	    0.4 & $\leq$0.2 & 569$\pm$15 &651  &     $>$3.5   \\ 
  Sz\,118                 & K5   &4067$\pm$ 82  & 4.47$\pm$0.37 &  $\leq$8.0   &$-$0.8$\pm$2.3 &	 0.5&	    0.5 & $\leq$0.2 & 558$\pm$10 &671  &2.95$\pm$0.13	\\ 
2MASS\,J16100133-3906449  & M6.5 &3018$\pm$ 72  & 3.77$\pm$0.23 & 15.9$\pm$5.4 &   0.1$\pm$2.6 &	 ...& $\leq$0.2 & $\leq$0.2 & 550$\pm$50 &521  &2.53$\pm$0.43	\\ 
SSTc2d\,J161018.6-383613  & M5   &3012$\pm$ 48  & 3.98$\pm$0.10 &  $\leq$8.0   &$-$0.2$\pm$2.5 &	 ...& $\leq$0.2 & $\leq$0.2 & 570$\pm$15 &541  &2.69$\pm$0.15	\\ 
SSTc2d\,J161019.8-383607  & M6.5 &2861$\pm$ 69  & 4.00$\pm$0.10 & 25.0$\pm$7.0 &   1.8$\pm$2.5 &      ...& $\leq$0.2 & $\leq$0.2 & 660$\pm$50 &630  &3.10$\pm$0.17   \\ 
SSTc2d\,J161029.6-392215$^a$&M4.5&3098$\pm$ 59 & 4.02$\pm$0.10 & 13.0$\pm$4.0 &$-$2.7$\pm$2.2 &      ...& $\leq$0.2 & $\leq$0.2 & 550$\pm$20 &522  &2.73$\pm$0.18   \\ 
SSTc2d\,J161243.8-381503  & M1   &3687$\pm$ 22  & 4.57$\pm$0.21 &  $\leq$8.0   &$-$2.3$\pm$2.4 &      ...&	   0.3 & $\leq$0.2 & 534$\pm$10 &585  &2.79$\pm$0.19  \\ 
SSTc2d\,J161344.1-373646  & M5   &3028$\pm$ 78  & 4.25$\pm$0.14 &  $\leq$8.0   &$-$1.2$\pm$2.3 &      ...&	   0.5 &       0.3 & 388$\pm$15 &503  &2.39$\pm$0.13  \\ 
\hline											        				      
\end{tabular}									        					      
\normalsize
\label{tab:param}
\end{table*}

\addtocounter{table}{-1}

\begin{table*}[htb]
\caption[]{{\it Continued.}}
\scriptsize
\begin{tabular}{llllrrrrrrrr}
\hline
\hline
\noalign{\smallskip}
 Name	                  & SpT & $T_{\rm eff}$ &   $\log g$   &  $v \sin i$  & $V_{\rm rad}$ & $r_{5400}$& $r_{6200}$& $r_{7100}$&$EW_{\rm Li}$ &$EW_{\rm Li}^{\rm corr}$&$A$(Li)  \\
                          &                            &    (K)       &    (dex)      &    (km/s)    &    (km/s)     &	     &  	 &	     & (m\AA)	   & (m\AA) & (dex) \\
\hline
\noalign{\smallskip}
\multicolumn{12}{c}{Sample of class II YSOs from the ESO Archive}\\
\hline
  GQ\,Lup		  & K6 &4192$\pm$ 65  & 4.12$\pm$0.36 &  $\leq$6.0   &   4.9$\pm$1.3 &     0.8 &	  0.5 &       0.3 & 483$\pm$10 &647  &3.17$\pm$0.10 \\
  Sz\,76$^a$              & M4 &3440$\pm$ 60  & 4.41$\pm$0.26 &  $\leq$6.0   &   1.4$\pm$1.0 &     0.7 & $\leq$0.2 & $\leq$0.2 & 609$\pm$50 &583  &  $>$3.5	 \\
  Sz\,77                  & K7 &4131$\pm$ 48  & 4.28$\pm$0.31 &  6.7$\pm$1.5 &   2.4$\pm$1.5 &     0.5 &	  0.4 & $\leq$0.2 & 573$\pm$15 &663  &3.08$\pm$0.12 \\
RX\,J1556.1-3655          & M1 &3770$\pm$ 70  & 4.75$\pm$0.21 & 12.7$\pm$1.3 &   2.6$\pm$1.2 &     ... &	  1.4 &       0.3 & 388$\pm$20 &671  &  $<$3.02     \\
  IM\,Lup$^a$	          & K5 &4146$\pm$ 95  & 3.80$\pm$0.37 & 17.1$\pm$1.4 &$-$0.5$\pm$1.3 &$\leq$0.2& $\leq$0.2 & $\leq$0.2 & 566$\pm$10 &545  &2.91$\pm$0.20 \\
  EX\,Lup                 & M0 &3859$\pm$ 62  & 4.31$\pm$0.29 &  7.5$\pm$1.9 &   1.9$\pm$1.4 &     1.0 &	  0.7 & $\leq$0.2 & 500$\pm$15 &642  &3.06$\pm$0.10 \\
~\\
\multicolumn{12}{c}{Sample of class III YSOs by \cite{frascaetal2017} observed within the GTO by \cite{alcalaetal2011}}\\
\hline
  Sz\,94		& M4   &3205$\pm$ 59  & 4.21$\pm$0.24 & 38.0$\pm$3.0 &  7.8$\pm$2.0  &     ... & $\leq$0.2 & $\leq$0.2 &   $<$28$^e$&$<$1 &  $< -1$	  \\ 
  Par-Lup3-1	        & M6.5 &2766$\pm$ 77  & 3.48$\pm$0.11 & 31.0$\pm$7.0 &  4.9$\pm$3.9  &     ... & $\leq$0.2 & $\leq$0.2 & 685$\pm$80 & 654 &  $<$3.25    \\
  Par-Lup3-2	        & M5   &3060$\pm$ 61  & 3.91$\pm$0.24 & 33.0$\pm$5.0 &  5.0$\pm$2.5  &     ... & $\leq$0.2 & $\leq$0.2 & 602$\pm$35 & 574 &3.09$\pm$0.18\\
  Sz\,107		& M5.5 &2928$\pm$100  & 3.52$\pm$0.13 & 66.6$\pm$4.1 &  6.1$\pm$3.8  &     ... & $\leq$0.2 & $\leq$0.2 & 570$\pm$30 & 540 &2.52$\pm$0.25\\
  Sz\,108\,A	        & ...  &3676$\pm$ 34  & 4.50$\pm$0.21 &  $\leq$8.0   &  0.4$\pm$2.2  &     0.5 &	 0.4 & $\leq$0.2 & 628$\pm$10 & 723 &3.02$\pm$0.04\\
  Sz\,121	        & M4   &3178$\pm$ 61  & 4.08$\pm$0.12 & 87.0$\pm$8.0 & 14.1$\pm$4.4  &     ... & $\leq$0.2 & $\leq$0.2 & 635$\pm$25 & 607 &  $>$3.50     \\
  Sz\,122		& M2   &3511$\pm$ 27  & 4.58$\pm$0.21 &149.2$\pm$1.0 & 17.9$\pm$3.0  &     ... & $\leq$0.2 & $\leq$0.2 & 300$\pm$25 & 274 &0.98$\pm$0.22\\
\hline		
\end{tabular}
~\\										      
$^a$ YSO with transitional disk; $^b$ Sub-luminous YSO; $^c$ Flat source (see \citealt{alcalaetal2017}); $^d$ Lithium non detected because of low $S/N$; 
$^e$ star with no lithium (see text for details). \\
Notes: Spectral types for class II and III YSOs were taken from \cite{alcalaetal2017} and \cite{manaraetal2013}, respectively, while effective 
temperatures, surface gravities, projected rotational velocities, radial velocities, and veiling in three spectral regions from \cite{frascaetal2017}.
\normalsize
\\
\end{table*}

\begin{table}[h]
\caption{Iron, barium, and lithium abundances measured through spectral synthesis. Errors refer to 
the best-fit procedure ($\sigma_{1}$), the main source of uncertainty (see text for details).} 
\label{tab:elemental_abundances}
\small
\begin{center}
\begin{tabular}{lccc}
\hline
\hline
\noalign{\smallskip}
Name &   [Fe/H] & [Ba/H] & $A({\rm Li})^{\rm synth}$ \\ 
     &   (dex)  & (dex) & (dex) \\ 
\hline
SSTc2d\,J160830.7-382827 &  $0.00\pm0.20$ & $0.75\pm0.20$ & $3.50\pm0.15$ \\
            RY\,Lup      &  $0.00\pm0.25$ & $1.10\pm0.30$ & $3.45\pm0.20$ \\
            MY\,Lup      &  $0.00\pm0.20$ & $0.70\pm0.25$ & $3.50\pm0.15$ \\
             Sz\,68      &  $0.10\pm0.30$ & $1.10\pm0.10$ & $3.90\pm0.20$ \\
            Sz\,133      &  $0.10\pm0.70$ & $1.20\pm0.70$ & $4.00\pm0.20$ \\
SSTc2d\,J160836.2-392302 &  $0.00\pm0.30$ & $1.05\pm0.10$ & $3.25\pm0.15$ \\
\hline
\end{tabular}
\end{center}
\normalsize
\end{table}

\section{Results and discussion} 
\label{sec:results_discussion}
\subsection{Lithium abundance, comparison with models, and implications} 
\label{sec:lithium_depl}
\begin{figure} %%[t!]
\begin{center}
\includegraphics[width=9cm]{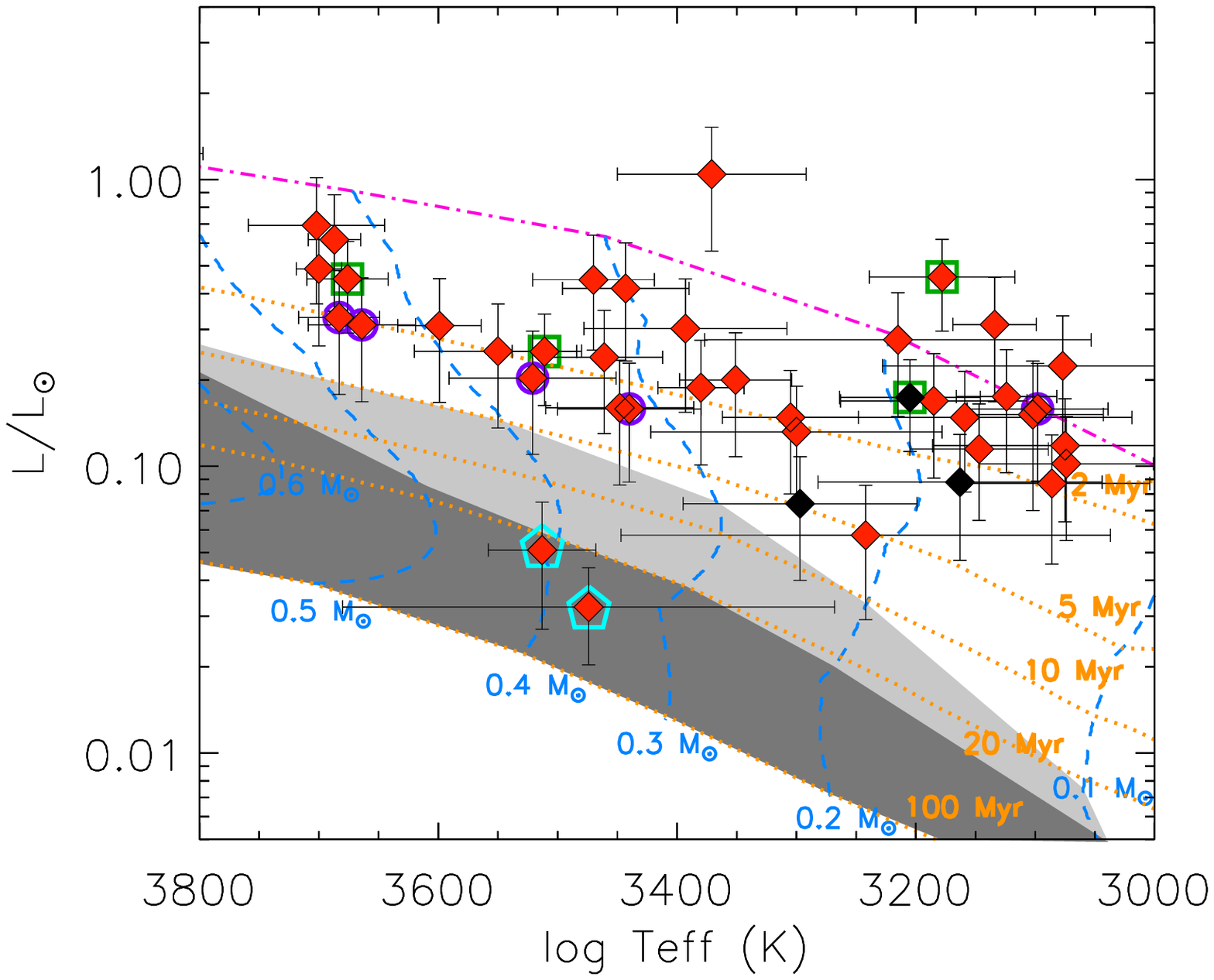}
\caption{Hertzsprung-Russell diagram for the sample stars with masses $\sim~0.2-0.5~M_\odot$. The shaded regions indicate different levels 
of predicted Li depletion: up to a factor of 10 (light gray) and more (dark gray) below the initial value and according 
to the mo\-dels of \cite{baraffeetal2015}. Masses from evolutionary tracks and isochronal ages are labeled. Black diamonds refer 
to targets showing depletion (i.e., Sz\,99, Sz\,69, and Sz\,94; see text and Fig.\,\ref{fig:AbunLi_Age}). Open squares, circles, 
and pentagons represent the position of the class III, transitional disk, and sub-luminous or flat targets, respectively.}
\label{fig:HRD} 
\end{center}
\end{figure}

Our sample comprises stars with $M_\star \sim 0.025-1.8 M_\odot$ (see \citealt{alcalaetal2017, frascaetal2017}). Considering the sub-sample of stars 
with masses $> 0.5 M_\odot$, we do not find an indication of Li depletion, $\sim 3.3$\,dex being the mean abundance of these stars. The situation is 
different for the sub-sample of stars with masses in the range $0.2-0.5 M_\odot$ (Fig.\,\ref{fig:HRD}). For these stars we can apply the so-called 
lithium test, as first proposed by \cite{pallaetal2005} for targets in Orion. Briefly, this is a useful clock based on the possibility that stars 
deplete their initial lithium content during the early phases of PMS contraction. It has been demonstrated that the theoretical assumptions 
required to study Li depletion hi\-sto\-ry have little uncertainty, as for fully convective stars the depletion mostly depends 
on $T_{\rm eff}$ (\citealt{bildstenetal1997}). At the same time, models show that stars with mass in the range $\sim 0.5-0.2 M_\odot$ start to 
deplete Li after 4--15 Myr and completely destroy it after a further 10--15 Myr (e.g., \citealt{baraffeetal2015}, and references therein). 
We therefore can compare the nuclear ages derived from the lithium test with the isochronal ages derived from the Hertzsprung-Russell (HR) diagram. 
First evi\-den\-ce of a Li depletion boundary for PMS stars have been reported by \cite{songetal2002} and \cite{whitehillenbrand2005}, where the 
timescale of Li depletion turned out to be larger than the isochronal age.

\begin{figure} %%[t!]
\begin{center}
\includegraphics[width=9cm]{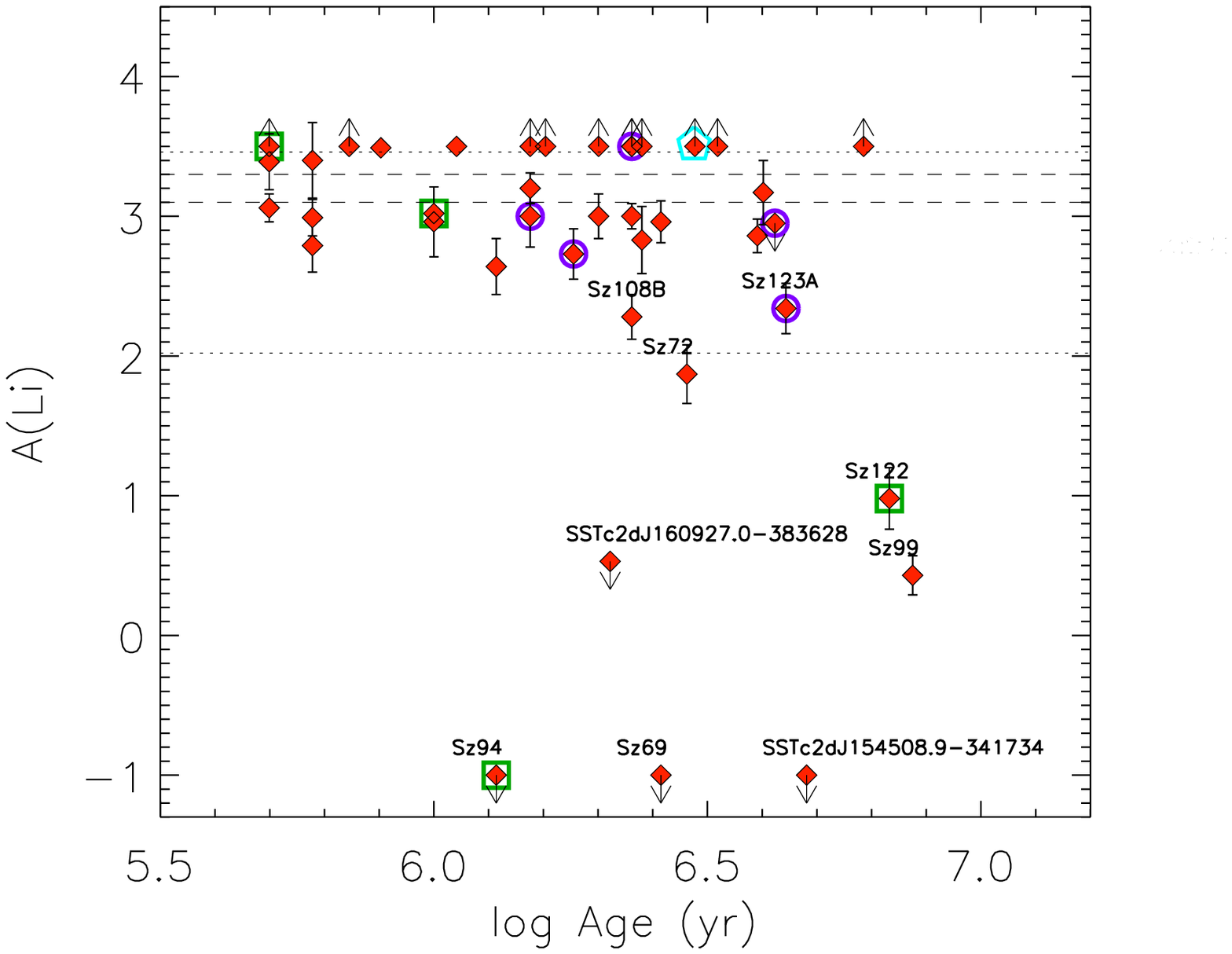}
\caption{Li abundance versus age for stars with mass in the range $0.2-0.5 M_\odot$. Symbols are the 
same as in Fig.\,\ref{fig:AbunLi_Teff}. The dashed horizontal lines mark the region of the interstellar lithium abundance (3.1--3.3 dex). 
Dotted lines represent $\pm 1 \sigma$ from the mean lithium abundance of this sub-sample.  
The labeled stars are those discussed in the text.}
\label{fig:AbunLi_Age} 
\end{center}
\end{figure}

In order to investigate the lithium depletion in our sample, we applied 
the procedure already tested in some sub-groups of the Orion SFR (see \citealt{pallaetal2005, pallaetal2007, saccoetal2007}) for objects with 
mass in the range $0.2 < M_\star < 0.5 M_\odot$. The derived lithium abundances are displayed in Fig.\,\ref{fig:AbunLi_Age} 
versus the age derived by \cite{frascaetal2017} using the \cite{baraffeetal2015} isochrones. The fi\-gu\-re 
shows a trend for the youngest stars to preserve their initial lithium content. Excluding the stars with upper and lower limits, the 
mean $A$(Li) for the sub-sample is around 2.74\,dex with a standard deviation of 0.72\,dex. Therefore, we consider as probable Li depleted 
targets those lying below a threshold of $\sim$2\,dex in lithium abundance.

\begin{figure} %[t!]
\begin{center}
\includegraphics[width=9cm]{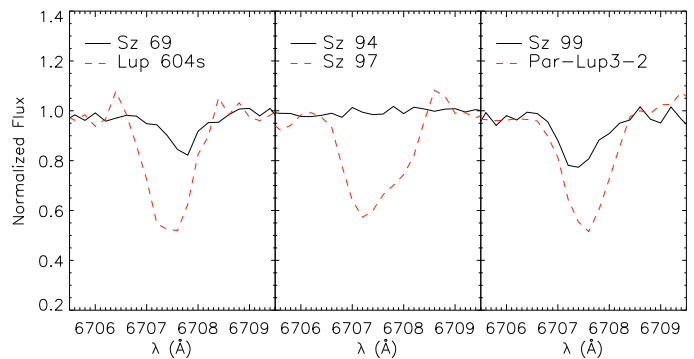}
\caption{Portions of spectra for the three Li depleted stars (solid lines) around the lithium line at $\lambda = 6707.8$\,\AA. 
Dashed lines represent the spectra of three non-Li depleted stars with similar stellar parameters (i.e., $T_{\rm eff}$, $\log g$, $v\sin i$, 
and $r_{6700}$).}
\label{fig:plot_spectra_Li} 
\end{center}
\end{figure}

Considering the stars with isochronal ages $> 2$\,Myr, five class II objects (SSTc2dJ160927.0-383628, Sz\,69, 
SSTc2dJ154508.9-341734, Sz\,72, and Sz\,99) and one class III (Sz\,122) appear below the threshold we defined, while two other class II YSOs 
(Sz\,108B and Sz\,123A) are close to it. However, slightly higher values of veiling, still within the uncertainty of $\sim 0.1-0.2$, 
could influence their abundance determination, placing them close to the non-depleted YSOs. Among the six apparently depleted 
targets, Sz\,72 has an effective temperature of 3550\,K, that is, in the $T_{\rm eff}$ within the range of overlap between the two 
curves of growth used to derive abundance (see Sect.\,\ref{sec:lithium_abun}), hence its $A$(Li) is rather uncertain. Then, for SSTc2d\,J160927.0-383628 
and SSTc2d\,J154508.9-341734 a veiling of $r_{6700} \sim 1$ is measured, hence their lithium abundances may be strongly influenced by strong continuum 
excess emission, while the class III star Sz\,122 could be a spectroscopic binary, as stressed in Sect.\,\ref{sec:lithium_abun}. 
In summary, among the six YSOs below our lithium depletion threshold, there are only two cases in which lithium depletion 
can be clearly assessed; these are Sz\,99 and Sz\,69. We thus consider these two YSOs as the most probable lithium depleted targets in our sample. 
In addition, we also include Sz\,94, a class III object for which lithium has not been detected either by us or other authors 
(\citealt{mortieretal2011, manaraetal2013, stelzeretal2013}), but its radial velocity and the proper motion reported by \cite{lopezmartietal2011} 
are consistent with the Lupus SFR. The profile of the \ion{Li}{i} $\lambda$6707.8\,\AA\,line 
for these three YSOs is shown in Fig.\,\ref{fig:plot_spectra_Li} together with the spectra of three targets with similar stellar parameters. 
The figure clearly shows the much weaker (or absent) lithium line for the three most probable Li-depleted stars, when compared with 
undepleted YSOs of similar $T_{\rm eff}$, $\log g$, $v \sin i$, and veiling.

\begin{figure}  %%[t!]
\begin{center}
\includegraphics[width=8.5cm]{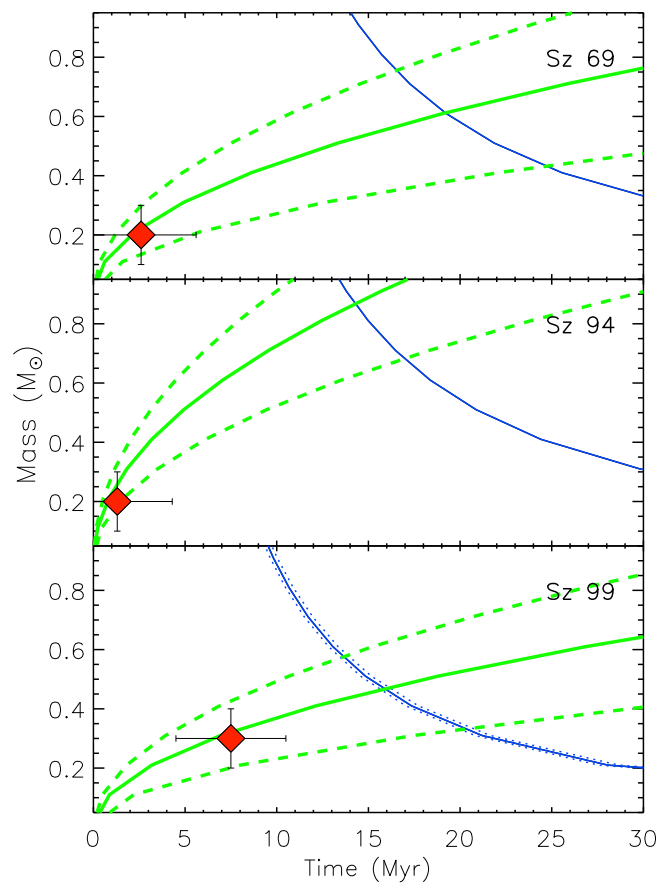}
\caption{Mass versus age for the three stars with evidence of Li depletion. The curves at a given luminosity (solid green line with positive slope) and 
at a given lithium abundance (solid blue line with negative slope) were computed for the values of $T_{\rm eff}$, luminosity, and lithium depletion given in 
Table\,\ref{tab:lithium_depleted_stars}. The dashed and dotted curves represent the uncertainty ranges in the observed luminosity and in the measured abundance, 
respectively. They define the locus where the values of mass and time are consistent with the observations. 
For Sz\,69 and Sz\,94 no errors in Li abundances are reported because their values, consistent with zero, are upper limits.
The diamonds give the mass and age from theoretical PMS tracks and isochrones, with typical errors of 
$\sim 0.1$\,$M_\odot$ and $\sim 3$ Myr, respectively (see \citealt{frascaetal2017}).}
\label{fig:MassTime_LiDepl} 
\end{center}
\end{figure}

The existence of very young Li-poor low-mass stars in Lupus allowed us to perform a comparison between theoretical and analytical models of early nuclear 
burning. We therefore adopted the stellar masses and isochronal ages determined by \cite{frascaetal2017} based on the PMS 
mo\-dels of \cite{baraffeetal2015}, and applied the procedure described in \cite{pallaetal2005}. Errors on both quantities were estimated 
from the uncertainties on luminosity and effective temperature provided by \cite{alcalaetal2017} and \cite{frascaetal2017}, respectively. 
As stressed by \cite{frascaetal2017}, individual ages can be affected by uncertainties depending on the data errors and the 
adopted set of evolutionary tracks. As such, these ages must be taken with care. We can compare these results with the analytic estimates 
derived by \cite{bildstenetal1997} for fully convective stars undergoing gravitational contraction at nearly constant $T_{\rm eff}$, assuming fast 
and complete mixing, and with negligible influence of degeneracy during the depletion, that is, in the mass range 
$0.2 M_\odot \ltsim M_\star \ltsim 0.5 M_\odot$. \cite{bildstenetal1997} derived relations for the time variation of the luminosity (see their Eq.\,4) 
and of the amount of Li depletion (see their Eq.\,11). Taking into account the measured luminosity and effective temperature, we can apply those equations 
and obtain a mass-depletion time plot. The results for the three Li-depleted stars are shown in Fig.\,\ref{fig:MassTime_LiDepl}. The line with positive 
slope represents the mass-age relation for each star at given $T_{\rm eff}$ and $L_\star$. The line with negative slope is the mass-age relation for a fixed Li 
abundance equal to that measured in each target. The intersection of these two lines yields the combination of mass and age at given $T_{\rm eff}$, 
luminosity, and lithium depletion. The derived values of nuclear masses and ages ($M_{\rm Li}$, $t_{\rm Li}$), together with those obtained by 
\cite{frascaetal2017} through HR diagram ($M_{\rm HRD}$, $t_{\rm HRD}$) for the three stars, are listed in Table\,\ref{tab:lithium_depleted_stars}. The 
three stars are within the bounded region predicted by the \cite{bildstenetal1997} analysis, but both masses and ages from \cite{baraffeetal2015} 
models are discrepant when compared to the nuclear masses and ages. In fact, while the HR diagram indicates 
masses of $\sim 0.2-0.3$\,$M_\odot$ and ages of $\sim 1-7$\,Myr, the amount of lithium depletion can only be explained by more massive ($\sim 0.5$\,$M_\odot$) 
and older ($\sim 15$\,Myr) stars (see Fig.\,\ref{fig:MassTime_LiDepl}). 
In all cases, the derived values of lithium abundance yield ages that are inconsistent with the isochronal ones: these targets have experienced 
too much burning for the estimated isochronal ages. In order to reconcile the two estimates, the stellar luminosity should be decreased by a 
factor of $\sim 2-5$ and/or the effective temperature increased by se\-ve\-ral hundreds of Kelvin (see Fig.\,\ref{fig:HRD}), which is several times 
more than the errors in these quantities. As pointed out by several authors (see \citealt{hartmann2003, pallaetal2007}, and references therein), 
the discrepancies may be explained as being due to a combination of effects, such as poorly known stellar (e.g., binarity, photometric variability, 
atmospheric parameters) and cluster (e.g., distance, extinction) properties. At the same time, an increase of the Li abundance up to the 
interstellar value would require an initial equivalent width a factor of $\sim 3-10$ larger than measured, which seems inconsistent if we take 
into account the uncertainties estimated in Section\,\ref{sec:lithium_abun}. Moreover, the influence of veiling in 
the $EW_{\rm Li}$ measurements is excluded because the three targets show $r_{6700}$ close to zero. In particular, to 
obtain $A$(Li)$\sim 3$ these targets should have a veiling of at least $\sim$3, which is not possible given their low level of 
continuum excess emission (see \citealt{alcalaetal2014, alcalaetal2017}). 

\begin{figure} 
\begin{center}
\includegraphics[width=9cm]{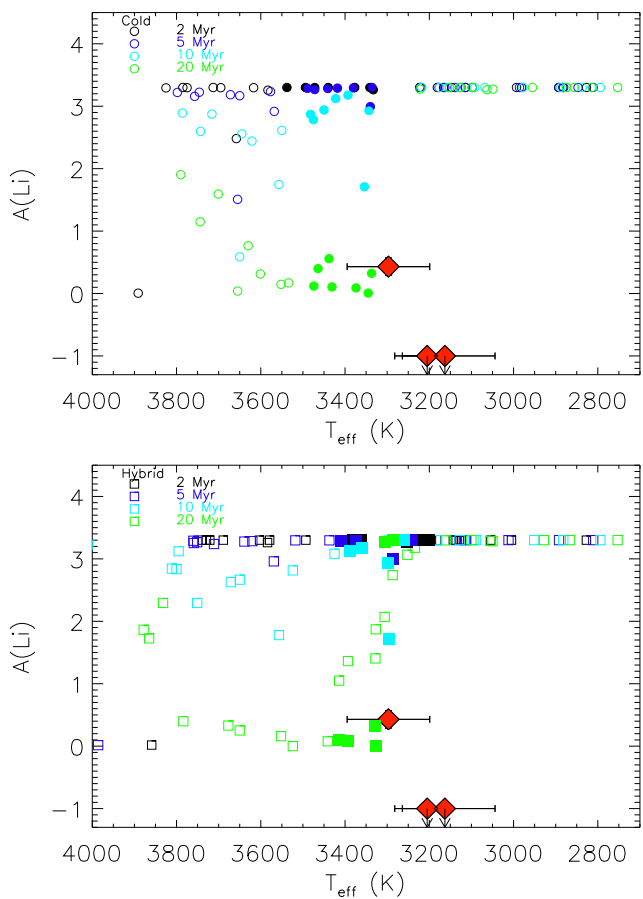}
\caption{Surface lithium abundance as a function of effective temperature in accreting models under the cold (upper panel) and hybrid 
(bottom panel) accretion scenarios computed by \cite{baraffeetal2017}. Black, blue, light blue, and green symbols refer to predictions computed at 
2, 5, 10, and 20\,Myr, respectively. Filled symbols refer to models that predict a final stellar mass close to that of our Li depleted 
targets (i.e., $\sim 0.2-0.3 M_\odot$). The positions of three Li depleted targets are shown with big red diamonds.}
\label{fig:AbunLi_Teff_Lidepl_Baraffeetal2017} 
\end{center}
\end{figure}

We also compared our results with recent self-consistent calculations coupling numerical hydrodynamics simulations of 
collapsing pre-stellar cores and stellar evolution mo\-dels of accreting objects (\citealt{baraffeetal2017}). These 
mo\-dels predict that early accretion of material with low internal 
energy (`cold accretion') or early accretion of material with energy depending on the accretion rate (`hybrid accretion') can produce objects 
with abnormal Li depletion, at odds with predictions from non-accreting stellar evolution models (\citealt{baraffeetal2017}, and references 
therein). In particular, accretion bursts with typical accretion rates $\dot M_{\rm burst} > 10^{-4} M_{\odot}$ yr$^{-1}$ may gravitationally 
compress the star, increasing its core temperature and pressure and triggering the early onset of lithium burning within a few Myr. Efficient 
large-scale convection, such as in low-mass PMS stars, would then rapidly deplete lithium throughout the star. Figure 
\ref{fig:AbunLi_Teff_Lidepl_Baraffeetal2017} shows the surface lithium abundances as a function of effective temperature in accreting models 
by \cite{baraffeetal2017} under cold and hybrid accretion scenarios for ages at 2, 5, 10, and 20 Myr (the median age of our data was estimated to 
be $\sim 2$ Myr; \citealt{frascaetal2017}). Our Li depleted targets show a content of lithium which is not reproduced by 
early accretion models. The latter predict less lithium depletion at the ages and masses (effective temperatures) of our targets. 
Only for one source do the mo\-dels seem to reproduce the observed lithium abundance but for an age ($\sim 20$\,Myr) inconsistent with the Lupus clouds.
We conclude that early accretion models are not able to reproduce the lithium depletion of such a rare object (3/89, i.e. $\sim 3\%$).

Therefore, we are not able to draw final conclusions about the origin of the lithium depletion in the three targets. 
We support the recent suggestion by \cite{baraffeetal2017} to devote more observational effort to characterize objects with abnormal Li depletion 
in young clusters, normally considered as outliers and often rejected from analyses. 

\begin{table*}[h]
\caption{Lithium depleted stars and their properties.} 
\label{tab:lithium_depleted_stars}
\begin{center}
\begin{tabular}{lccccccr}
\hline
\hline
\noalign{\smallskip}
Name & $L_\star$ & $T_{\rm eff}$ & $A$(Li) & $M_{\rm HRD}$ & $t_{\rm HRD}$ &  $M_{\rm Li}$ & $t_{\rm Li}$  \\ 
    & ($L_\odot$) & (K) & (dex) & ($M_\odot$) & (Myr) &  ($M_\odot$) & (Myr) \\ 
\hline
Sz\,69  & $0.088\pm0.041$ & $3163\pm119$& $< -1$        & $0.2$ & $2.6$ & $\sim 0.6$ & $\sim 19$\\
Sz\,94  & $0.174\pm0.061$ & $3205\pm59$ & $< -1$        & $0.2$ & $1.3$ & $\sim 0.9$ & $\sim 14$\\
Sz\,99  & $0.074\pm0.034$ & $3297\pm98$ & $0.43\pm0.14$ & $0.3$ & $7.5$ & $\sim 0.5$ & $\sim 16$\\
\hline
\end{tabular}
\end{center}
\normalsize
\end{table*}

\subsection{Iron abundance and comparison with previous works}
Only one of our targets (namely, RY\,Lup) is in common with previous studies of iron abundance in the literature. For 
this star, we obtain a best-fit at [Fe/H]=$0.00\pm0.25$, which is in agreement with the value of [Fe/H]=$-0.09\pm0.13$ 
derived by \cite{padgett1996}.

\begin{figure} 
\begin{center}
\includegraphics[width=9cm]{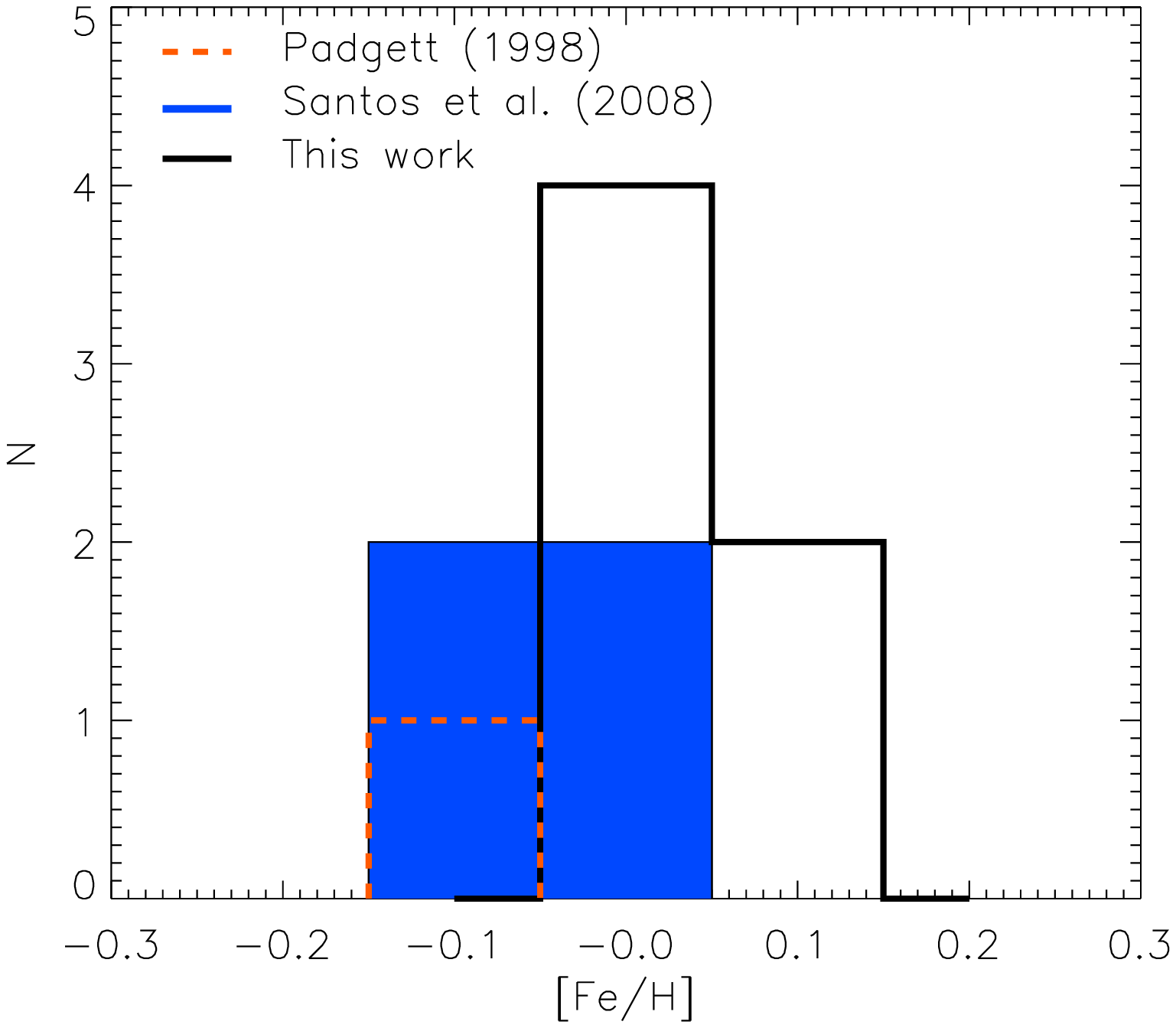}
\caption{Comparison of our [Fe/H] with previous estimates by \cite{padgett1996} and \cite{santosetal2008}. }
\label{fig:fe_distr} 
\end{center}
\end{figure}

In Fig.\,\ref{fig:fe_distr} the distribution of our [Fe/H] measurements in Lupus is compared with the previous estimates by 
\cite{padgett1996} and \cite{santosetal2008}. In the first case, only the abundance of RY\,Lup was measured. In the se\-cond work, the average value 
obtained by analyzing four class III stars was <[Fe/H]>=$-0.05\pm0.01$, which is in agreement with our estimate. We note that, 
even if our measurements have larger uncertainties when compared with previous works (mainly because of the relatively low 
spectral resolution of the instrument used here), our distribution is narrow, peaked at $\sim 0.00$\,dex with a standard deviation of 0.05\,dex. 

We stress that our abundance determination on class II targets is one of the few in which the contribution of veiling has been taken into account. 
A previous paper providing abundance determination for ``deveiled'' class II targets is that by \cite{dorazietal2011} focused on the Taurus-Auriga SFR, 
where the authors find that class II and class III objects share the same chemical composition, in\-di\-ca\-ting that 
the presence of a circumstellar accretion disk does not affect the stellar photospheric abundances. Other older analyses in class II targets 
did not consider the correction for the veiling, claiming that the derived iron abundance had to be treated as a lower limit 
(see, e.g., \citealt{padgett1996}). 

\subsection{Iron abundance in the context of young stellar clusters}
\label{sec:ironabundance_youngclusters}

Recent works have shown that nearby ($<500$\,pc) young open clusters span a range in [Fe/H] from $\sim -0.20$ to $\sim +0.30$ dex, but the youngest 
associations ($\ltsim 100$\,Myr) are generally clustered around the low metallicity values (see Fig.\,13 in \citealt{biazzoetal2011a} and Fig.\,10 in 
\citealt{spinaetal2014b}). Because of their young ages, these regions did not have time to migrate through the Galactic disk. Therefore, their 
metal content should be representative of the present chemical pattern of the nearby interstellar gas from which their members formed, with negligible 
effects of chemical evolution (\citealt{biazzoetal2011a}, and re\-fe\-ren\-ces therein). \cite{spinaetal2014b} concluded that since the chemical content provides 
a powerful tool for tagging groups of stars to a common formation site, these dif\-fe\-rent young regions should share the same origin. They also 
report the metallicity distribution of regions younger than 100\,Myr, separating clusters associated and not associated with the Gould Belt (see their 
Fig.\,11), a disk-like structure made up of gas, young stars, and associations, whose origin remains somewhat controversial (see, e.g., 
\citealt{guilloutetal1998, bally2008}, and references therein). \cite{spinaetal2014b} also claim that star-forming regions and young open clusters 
asso\-cia\-ted with the Gould Belt show a metallicity lower than the solar value, and this could be the reason for the ``non metal-rich'' nature 
of the youngest stars in the solar vicinity. 

Our determination of iron abundance for the Lupus targets with the X-shooter spectrograph is not as accurate as in the cited 
works, mainly because of the relatively low re\-so\-lu\-tion of our spectra, but, within the errors, it is in line with recent results. However, 
we caution about the conclusion related to the Gould Belt for a number of reasons: the number statistics of the YSOs in 
the regions studied so far is still too low, the abundance determinations are rather heterogeneous because of the different methodologies used 
to derive them, and the uncertainties in the distance and proper motion of the studied YSOs are still rather high. 
Therefore, additional accurate and homogeneous determinations of elemental abundance and kinematics are needed to further investigate this issue. 
Present and future facilities both from space (e.g., {\it Gaia}) and from the Earth (e.g., the {\it Gaia-ESO Survey}) will allow us to trace dynamically and 
chemically our Galactic disk with unprecedented accuracy.

\subsection{The barium problem}
\label{sec:barium_issue}
\begin{figure} 
\begin{center}
\includegraphics[width=9cm]{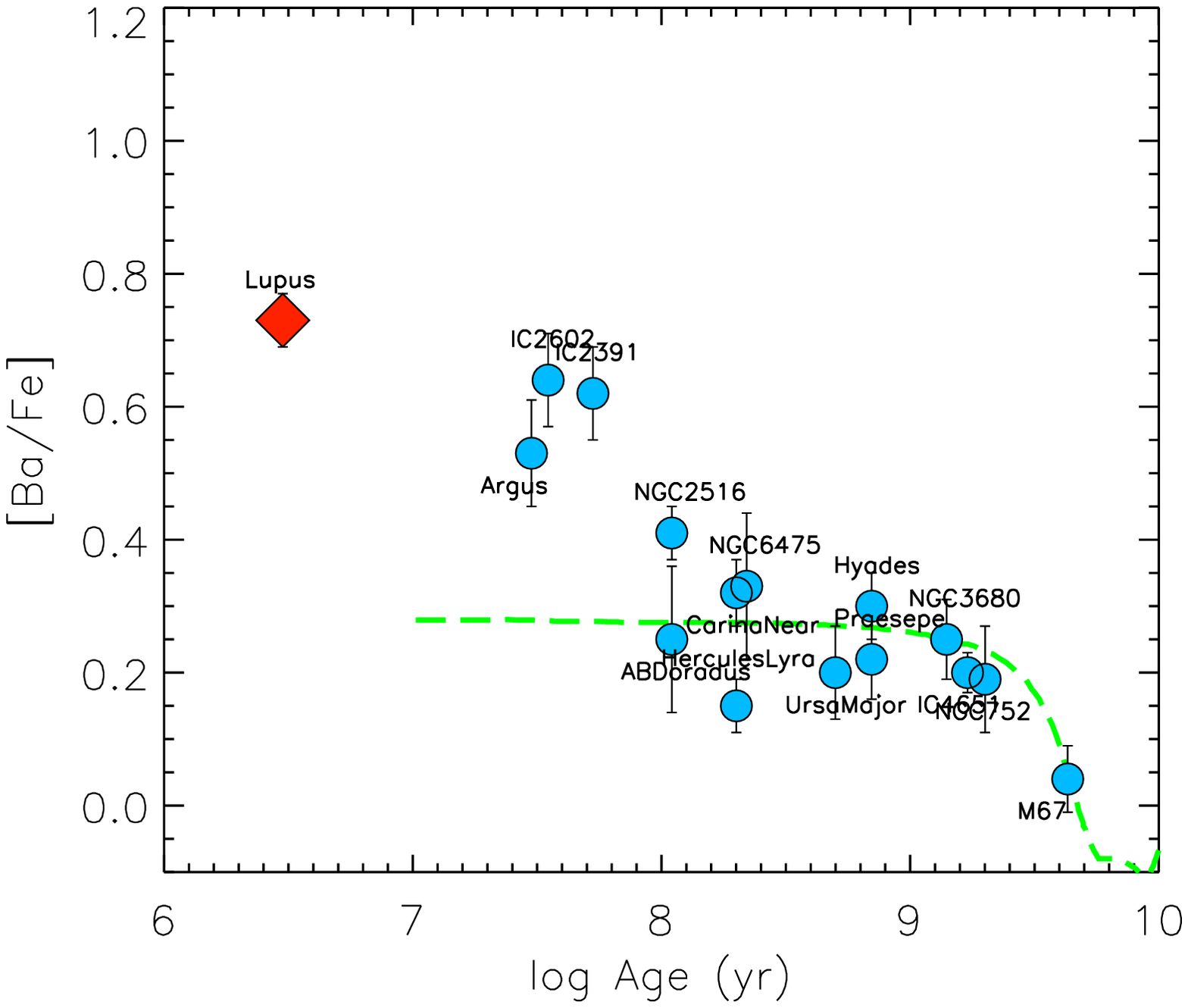}
\caption{[Ba/Fe] abundance as a function of cluster age for dwarf stars in Lupus (this work) and other regions (from the 
literature). The red diamond represents the mean value of the most reliable Ba abundances obtained for the two warmest targets (see text and 
Table\,\ref{tab:elemental_abundances}). Cluster abundances and ages were taken from \cite{dorazietal2009, dorazietal2012}, \cite{desilvaetal2013}, 
and \cite{reddylambert2015}. The dashed line represents the Galactic chemical evolution model developed by \cite{dorazietal2009} adopting enhanced $s$-process 
yields from AGB stars.}
\label{fig:ba_age} 
\end{center}
\end{figure}

In Fig.\,\ref{fig:ba_age} we show the [Ba/Fe] ratio versus mean age for targets in clusters studied by several authors 
(\citealt{dorazietal2009, dorazietal2012, desilvaetal2013, reddylambert2015}). For homogeneity reasons, we considered only dwarf targets 
analyzed with similar methods to ours. The Ba abundance shows an in\-crea\-si\-ng trend with decreasing age, as already reported 
in previous works (e.g., \citealt{dorazietal2009, desilvaetal2013, yongetal2012, jacobsonfriel2013}). To our knowledge, the sample of 
dwarfs in Lupus is the youngest one analyzed so far for barium abundance determinations. 

\begin{figure} 
\begin{center}
\includegraphics[width=9cm]{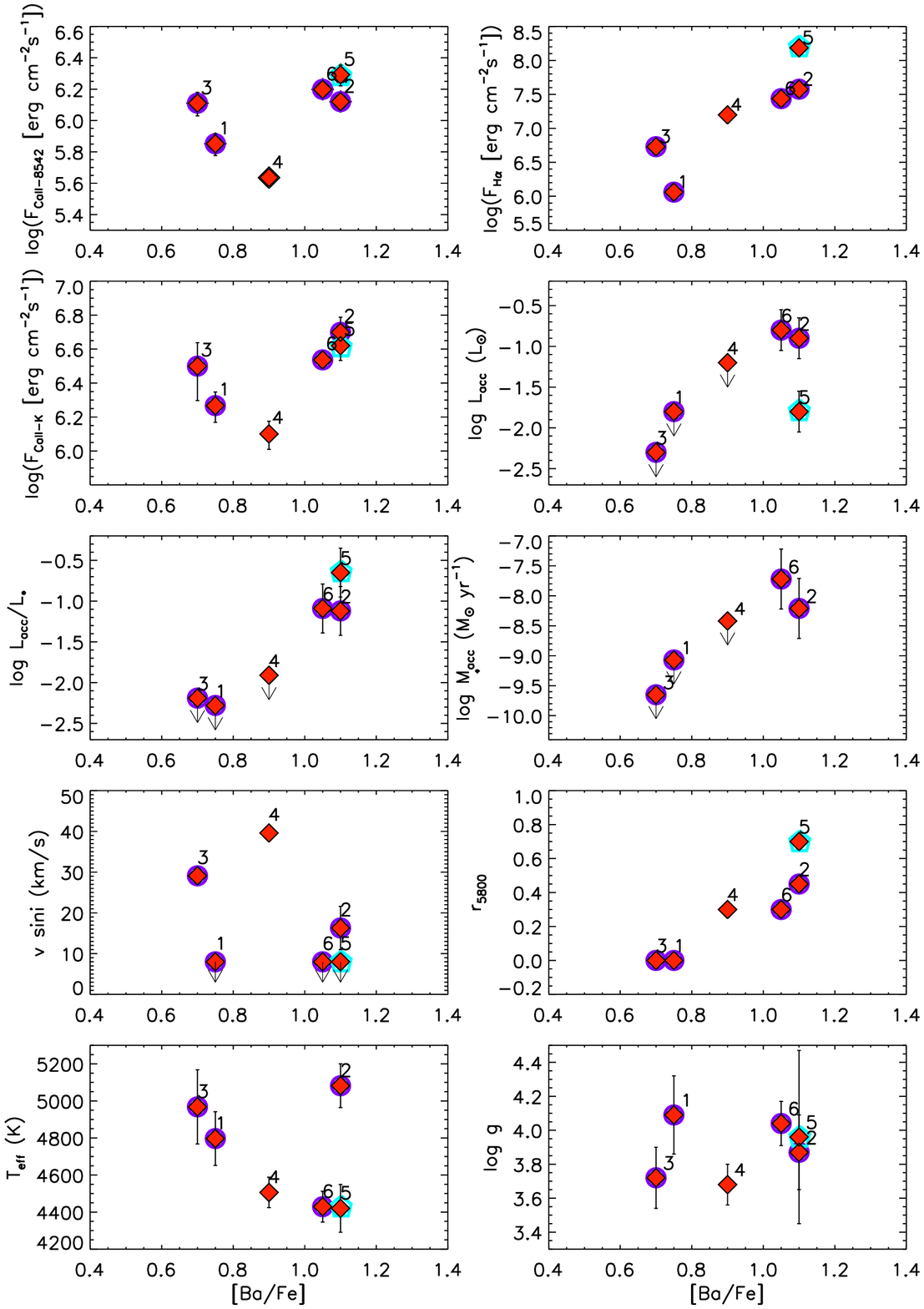}
\vspace{.5cm}
\caption{Dependence of the [Ba/Fe] abundance on activity (\ion{Ca}{ii}-8542, H$\alpha$, \ion{Ca}{ii}-K fluxes) and accretion ($L_{\rm acc}$, 
$L_{\rm acc}/L_\star$, $\dot M_{\rm acc}$, $r_{5800}$) diagnostics and on stellar parameters ($v \sin i$, $T_{\rm eff}$, $\log g$). Stars are 
enumerated: SSTc2d\,J160830.7-382827 (1), RY\,Lup (2), MY\,Lup (3), Sz\,68 (4), Sz\,133 (5), and SSTc2d\,J160836.2-392302 (6). 
Open circles and pentagons represent transitional disk and sub-luminous targets, respectively. Error bars of the [Ba/Fe] abundances 
are not plotted for clarity reasons.}
\label{fig:ba_activ_accr} 
\end{center}
\end{figure}

For targets older than $\sim$500 Myr, the trend of increa\-sing [Ba/Fe] with decreasing cluster age was interpreted by \cite{dorazietal2009} 
as an indication that low-mass ($<$~$1.5\,M_\odot$) AGB stars contributed more importantly to the chemical evolution of the Galactic disk than 
previously assumed by models (see Fig.\,\ref{fig:ba_age}). Applying $s$-process yields in which the neutron source was larger by a factor six 
than in previous models, \cite{dorazietal2009} could reproduce the observed trend (see also \citealt{maiorcaetal2014}). 
On the other hand, as the same authors acknowledge, it would be difficult to imagine a process capable of creating barium in the 
last 500 Myr of Galactic evolution, unless local enrichment is assumed. Thus, in agreement with other 
works analyzing dwarf stars in 30-50\,Myr old associations and clusters (e.g., \citealt{dorazietal2009, dorazietal2012, dorazietal2017}), we 
are still facing a barium conundrum. In our case, the behavior is even more extreme, because for the Lupus SFR our [Ba/Fe] values are 
$\gtsim 0.7$\,dex (Fig.\,\ref{fig:ba_age}). If, on one side, the sharp function with age is evident, on the other, whether this corresponds 
to a real increase in the Ba abundance or whether it depends on the methodologies used to derive the abundance is still a matter of debate (see also 
\citealt{dorazietal2017}). To spot possible artifacts in our analysis of Ba abundance, we checked for plausible 
dependence of [Ba/Fe] on stellar parameters, chromospheric activity levels, and accretion properties. This is 
depicted in Fig.\,\ref{fig:ba_activ_accr} and discussed next. 

At solar metallicity, the formation of the $\lambda$5853.7\,\AA\,\ion{Ba}{ii} line occurs in atmospheric layers that are mostly 
at effective depths below $\log \tau_{5000}=-1.9$ (\citealt{mashonkinazhao2006}), therefore quite deep to expect a strong 
impact from the above hot chromosphere. Indeed, we do not see any clear relationship between the activity indicators \ion{Ca}{ii} 
fluxes (or rotational velocity) derived by \cite{frascaetal2017} and Ba over-abundance. This is in agreement 
with the findings by \cite{dorazietal2012} for young clusters of $\sim$30-50\,Myr. A trend with the H$\alpha$ 
flux seems to be instead present, maybe related to the different phy\-si\-cal conditions of the emitting regions in \ion{Ca}{ii} and H$\alpha$ lines (see 
\citealt{frascaetal2017}, and references therein). In fact, for the six stars, the \ion{Ca}{ii}-IRT flux ratio is around 
1.2-1.6, typical of optically thick emission sources, while the Balmer decrement is around 3-10, typical of optically thin emission (see 
\citealt{frascaetal2017} for details).

No $\log g$-[Ba/Fe] relation seems to be present, while some $T_{\rm eff}$-[Ba/Fe] trend appears, with the exception of one target (RY Lup). 
This trend could reflect some effect of overionization, with the consequence of abundance dif\-fe\-ren\-ces between the neutral species and the 
singly-ionized ones, as found for several iron-peak and $\alpha$- elements for dwarf targets in $\sim$30-250 Myr old clusters (see 
\citealt{schuleretal2003, schuleretal2004, dorazirandich2009}). This effect seems to be strong at temperatures cooler than $\sim$5000\,K, 
where the neutral element shows lower abundance with respect to the first ionized element. To confirm that overionization is the cause of 
the high \ion{Ba}{ii} abundances of cooler stars, we should also compute the Ba abundances using \ion{Ba}{i} lines. In our case, we could exploit 
the \ion{Ba}{i} 5535\,\AA\,line, as suggested by \cite{reddylambert2015}, but, as the same authors claim, this feature is 
strongly affected by departures from LTE; thus, until corrections are provided for this line, we are not able to 
use it as an alternative approach. As underlined in Sect.\,\ref{sec:barium_abun}, microturbulence and veiling 
have a strong impact on barium abundance determination. In any case, a barium abundance around the solar value can be reached only with changes in 
microturbulence and veiling by even more than 100\%, which is unrealistic. In conclusion, stellar parameters seem to be not (at least solely) responsible 
for the barium over-abundance.

Even if we are aware that we have at our disposal a very low number statistics, another trend that seems to appear in Fig.\,\ref{fig:ba_activ_accr} is 
between the Ba abundance and the accretion luminosity $L_{\rm acc}$ (and mass accretion rate $\dot M_{\rm acc}$) derived by \cite{alcalaetal2017}. 
Similar behavior is seen with the veiling, as weak accreting targets show also lower $r_{5800}$ va\-lues, and in our case smaller Ba abundance. We 
tentatively attribute this trend to the effects of strong, warm flux from the upper chromosphere due to 
strong accreting phenomena on the structure of the stellar atmosphere. 

The two targets which suffer less from all these effects are the warmest and ``unveiled'' SSTc2d\,J160830.7-382827 and MY\,Lup, both of them having transitional 
disks and being weak (or dubious) accretors, because their accretion luminosity is comparable to the chromospheric level (see \citealt{alcalaetal2017}). In 
summary, with the aim of avoiding problems related to possible dependence of [Ba/Fe] on accretion or activity diagnostics and stellar parameters, we consider as 
the most reliable those derived for the two warmest weak accretors (i.e., SSTc2d\,J160830.7-382827 and MY\,Lup), 
with an average [Ba/Fe]$\sim 0.7$ dex (red diamond in Fig.\,\ref{fig:ba_age}).

In conclusion, we are not able to provide an explanation for the peculiar trend of the Ba abundance in young SFRs, clusters, and associations. We 
also do not think that any of the examined possibilities can explain the observed Ba over-abundance at the $\sim$0.7\,dex level. Therefore, other 
phy\-si\-cal reasons must be at work. Recently, \cite{dorazietal2017} claimed the activation of the intermediate $i$-process as a 
promising mechanism of production of heavy elements (see also \citealt{misheninaetal2015, hampeletal2016}). Howe\-ver, 
further theoretical work is needed. Future observational surveys for the determination of Ba elemental 
abundance with homogeneous methodologies in large samples of stars in young clusters will be of paramount importance 
for good statistics at ages of $\ltsim 50$\,Myr.

\section{Conclusions}
\label{sec:conclusions}
We have presented the results of a study on elemental abundances in the Lupus star-forming region using 
spectroscopic data acquired with X-shooter at the VLT. The studied sample comprises of almost all class II objects in the 
Lupus I, II, III, and IV clouds (82 over 89 sources; see \citealt{alcalaetal2017}) and seven class III objects. Three elements 
were analyzed: lithium, iron, and barium.

Our main results can be summarized as follows:
\begin{itemize}

\item We detected the lithium line at $\lambda$=6707.8\,\AA\, in all targets but 
six. The class III object Sz\,94 does not show any hint of the absorption line, while for the other five class II targets we could only measure 
upper limits for the lithium equivalent widths due to the low $S/N$ of the spectra.
\item Three objects appear to be highly lithium depleted. They represent only a few percent of the Lupus population, hence they 
are extremely rare, as found in other star-forming regions. The depletion in the lithium elemental abundance observed in such objects is still not 
reproduced by pre-main sequence evo\-lu\-tio\-na\-ry models in the literature, which makes them appealing for future detailed studies. 
\item For six class II targets we measured iron and barium abundance through spectral synthesis. 
The mean iron abundance in the Lupus star-forming region is con\-si\-stent, within the errors, with the chemical pattern of 
the Galactic thin disk in the solar neighborhood.
\item We found enhancement in barium abundance up to $\sim 0.7$\,dex level. Our targets thus confirm and extend to a younger age 
that previously found by other authors. We discussed several possible explanations for this puzzling behavior, including 
chromospheric and accretion effects, uncertainties in stellar parameters, and departure from LTE approximation, but none of these seems 
to completely justify the barium over-abundance. The barium problem is still an open issue and deserves further work, both theoretical 
and observational, in particular at clu\-ster ages $\ltsim 50$\,Myr.

\end{itemize}

\begin{acknowledgements}
The authors are very grateful to the referee for his/her useful remarks that allowed us to improve the previous version of the manuscript. 
The authors wish to dedicate this paper in remembrance of Francesco Palla; in particular KB is grateful to Francesco for his enriching 
teaching in the field of star formation, his extraordinary simplicity and his humility. KB also thanks Valentina D'Orazi and Chris Sneden for 
fruitful discussions on the barium issue. CFM gratefully acknowledges an ESA Research Fellowship. This research has made use of the 
SIMBAD database, operated at CDS (Strasbourg, France). 
\end{acknowledgements}

\bibliographystyle{../../../aa-package_Feb2016/aa}

\end{document}